\documentclass[reprint,notitlepage,tightenlines,nofootinbib,eqsecnum,superscriptaddress,preprintnumbers,12pt]{revtex4}
\pdfoutput=1
\usepackage{amsmath,latexsym,amssymb,graphicx,ifpdf,slashed,color,hyperref,url}
\hypersetup{colorlinks,citecolor= nicegreen,linkcolor= nicered}
\definecolor{nicered}{rgb}{0.7,0.1,0.1}
\definecolor{nicegreen}{rgb}{0.1,0.5,0.1}
\newcommand{\beq}{\begin{equation}}
\newcommand{\eeq}{\end{equation}}
\newcommand{\beqa}{\begin{eqnarray}}
\newcommand{\eeqa}{\end{eqnarray}}
\newcommand{\bea}{\begin{eqnarray}}
\newcommand{\eea}{\end{eqnarray}}
\newcommand{\bi}{\begin{itemize}}
\newcommand{\ei}{\end{itemize}}
\newcommand{\ben}{\begin{enumerate}}
\newcommand{\een}{\end{enumerate}}

\def\Cornell{Department of Physics, LEPP, Cornell University, Ithaca, NY 14853, USA}
\def\Northwestern{Department of Physics and Astronomy, Northwestern University, Evanston, IL 60208, USA}
\def\Fermilab{Theoretical Physics Department, Fermilab, P.O. Box 500, Batavia, IL 60510, USA}

\begin{document}

\title{
Self-Destructing Dark Matter}

\author{Yuval Grossman}
\email[Electronic address: ]{yg73@cornell.edu} 
\affiliation{\Cornell}
\author{Roni Harnik}            
\email[Electronic address: ]{roni@fnal.gov}
\affiliation{\Fermilab}
\author{Ofri Telem}
\email[Electronic address: ]{t10ofrit@gmail.com} 
\affiliation{\Cornell}
\author{Yue Zhang\,}
\email[Electronic address: ]{yuezhang@northwestern.edu} 
\affiliation{\Fermilab}
\affiliation{\Northwestern}

\date{\today}
\begin{abstract}
We present Self-Destructing Dark Matter (SDDM), a new class of dark matter models which are detectable in large neutrino detectors. In this class of models, a component of dark matter can transition from a long-lived state to a short-lived one by scattering off of a nucleus or an electron in the Earth. The short-lived state then decays to Standard Model particles, generating a dark matter signal with a visible energy of order the dark matter mass rather than just its recoil.
This leads to striking signals in large detectors with high energy thresholds.
We present a few examples of models which exhibit self destruction, all inspired by bound state dynamics in the Standard Model. The models under consideration exhibit a rich phenomenology, possibly featuring events with one, two, or even three lepton pairs, each with a fixed invariant mass and a fixed energy, as well as non-trivial directional distributions. This motivates dedicated searches for dark matter in large underground detectors such as Super-K, Borexino, SNO+, and DUNE. 
\end{abstract} 

\preprint{FERMILAB-PUB-17-505-T}
\preprint{NUHEP-TH-17-07}

\maketitle

\section{Introduction}

Our knowledge of the nature of dark matter (DM) is very limited. Though it accounts for over 80\% of the matter in the universe, it is often assumed that the dark sector is minimal, consisting of a single particle that interacts only feebly with the Standard Model (SM). This assumption, which is the most sensible first guess, motivates a set of strategies in our search for DM. In particular, when searching for DM directly, we try to detect the low energy recoil of a nucleus which was struck by a slowly moving DM particle~\cite{Goodman:1984dc}. DM direct detection thus focuses on detectors with thresholds between sub keV and tens of keV.

The current search program is undoubtedly important and must be pursued further. However, we should not let minimality be our only guide. Indeed, considering even slightly non-minimal DM models has led to significant variations in direct detection phenomenology~\cite{TuckerSmith:2001hy, Finkbeiner:2007kk, Chang:2008gd, Bramante:2016rdh, Chang:2008xa, Graham:2010ca, Feldstein:2010su,Essig:2010ye, Bai:2009cd, Pospelov:2008qx, An:2012bs, Chang:2009yt,Chang:2010en,Pospelov:2013nea,Kumar:2011iy}, including changes in the predicted recoil spectrum, annual modulation, and in the manner in which the DM deposits energy in the detector.
These interesting models, however, keep direct detection phenomenology within the purview of low threshold detectors. The reason for this is that the available energy in the problem is at most the kinetic energy in the DM-nucleus or DM-electron system. This energy is of order $\mu v^2$, where $v\sim 10^{-3}c$ is the DM velocity, and $\mu$ is the reduced mass of the system. 

In this paper, we consider a drastic departure from this picture and discuss the possibility that DM is capable of leaving not just kinetic energy in the detector, but rather \emph{all} of its mass. This can lead to DM direct detection signals above MeV, allowing to extend the search for DM to detectors with much higher thresholds. In particular, large existing neutrino experiments such as Borexino and Super-Kamiokande, as well as future experiments such as SNO+ and DUNE, can probe self destructing DM models.

\begin{figure}[t]
\centerline{\hspace{2cm}\includegraphics[height=5cm]{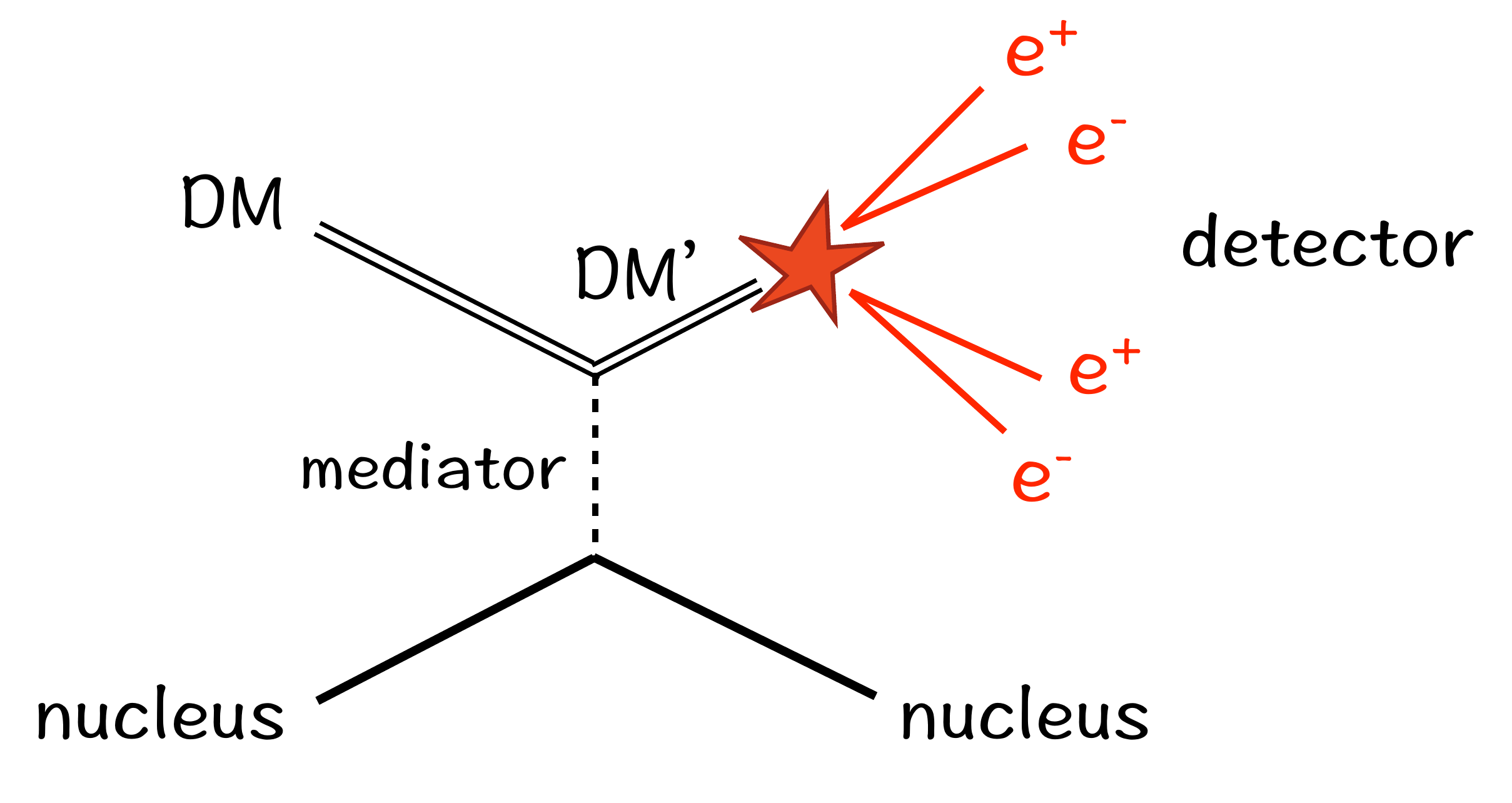}}
\caption{A schematic picture of self-destructing DM, where DM denotes the cosmologically long-lived state and DM' denotes the very short-lived one.}
\label{fig:cartoon}
\end{figure}

A na{\"i}ve idea to convert the DM rest mass to detectable signal is to consider its down-scattering to a nearly massless state, with the difference in mass converted to recoil energy. This is however unfeasible since it leads to a DM particle that is cosmologically unstable by decaying to this nearly massless state. Instead, we must consider a non-minimal setup in which an interaction of DM with a nucleus triggers a transition from a state that is stable on cosmological timescales to one which is very short-lived. The signal is then not a recoil signal, but rather a subsequent decay of the short-lived state to SM particles. A sketch of this is depicted in Figure~\ref{fig:cartoon}.
The crucial ingredients in the story are a cosmologically long lived state and a short lived state, with a possible transition between the two which is induced by the interaction with the Earth.  We call this class of DM models self-destructing DM (SDDM for short).

SDDM is not the first class of DM models in which the measured energy is more than just the DM recoil energy. In \cite{Graham:2010ca},
for example, the DM down-scatters and converts part of its rest mass to recoil energy. However, in that model the mass splittings considered were still of the same order as the kinetic energy. In~\cite{Pospelov:2013nea} the possibility of releasing more than the recoil energy was considered, but with signals still in the domain of low threshold detectors. Other DM models \cite{Davoudiasl:2011fj, Davoudiasl:2010am, Huang:2013xfa,Agashe:2014yua,Berger:2014sqa,Kong:2014mia,Alhazmi:2016qcs}, are similar to our scenario in that they can be probed at neutrino detectors~\cite{Kachulis:2017nci}. Yet, in these models, the dynamics that lead to signals at neutrino detector (a ``removal'' of a baryon from the detector or a boosted population of DM) is different than the case of SDDM. Lastly, we mention that the axion and other ultralight bosons that can be fully absorbed by a detector may be considered a fully SDDM candidate, but their lightness leads to much lower energy deposition, and thus to very different experimental signatures~\cite{Sikivie:1983ip, Chaudhuri:2014dla, Hochberg:2016sqx}.

The self-destructing scenario is possible thanks to the high density of matter in the Earth, one which is unprecedented from the perspective of an incoming DM particle.  The expected number of DM interactions off of a target with number density $n$ is roughly 
\begin{equation}
\langle N \rangle \sim n\langle\sigma v\rangle\Delta t\,, 
\end{equation}
where $v$ is the typical relative velocity, $\sigma$ is the interaction cross section, and $\Delta t$ is the amount of time spent in this environment.  A DM particle with $v \sim 10^{-3}c$ crosses the Earth in about 10 seconds. However, before its short traversal through Earth, it had spent about $10^{16}$ more time in the low density environment of the galaxy. On the other hand, the number density of atoms in the Earth is about $10^{23}$ larger than that in the galaxy.  As a result, it is reasonable for the expected rate of transitions to an unstable state to be much larger in one Earth-crossing than it is in the preceding Hubble time. Of course, one should not forget that in the early universe the DM candidate was immersed in a dense thermal bath (In particular, the baryon number density of the universe is comparable to that of the Earth at temperatures around MeV scale). Avoiding self-destruction of dark matter in these early times will be a requirement for our models and will be discussed later.

In the upcoming sections, we present three models which realize the idea of SDDM.  
These ideas are intended as a proof of principle rather than as complete, well motivated models. In each case, DM is stabilized by a different principle that can be un-done by interactions in the Earth.  None of these models is minimal by any means. They are, however, inspired by phenomena that are observed in the visible sector.  Particles in the Standard Model find themselves to be either short-lived or meta-stable for a variety of reasons.  In many cases different effects, forces, or interactions come into play, balancing one against the other to produce a long- or short-lived state.  We are therefore quite liberal in allowing for several particles and interactions.  Furthermore, like the visible sector which contains several populations of cosmic relics, we will be open to the possibility that only a subcomponent of DM is self-destructing.

The model building challenge for SDDM is to have a dark matter component that is very long-lived, yet capable self-annihilating once it interacts with a nucleus in the Earth. In its long-lived state, this component needs to have a lifetime for more than about $10^{28}$ sec to satisfy observational constraints~\cite{Arvanitaki:2008hq, Arvanitaki:2009yb, Essig:2013goa}, if it makes up all of DM. Even if we consider only a small sub-component of DM to relax this bound, the state needs to live for more than the age of the universe. In contrast, the short-lived state of our DM component has to decay in much less than 10 seconds, its Earth-crossing time.

The paper is organized as follows:
In Section~\ref{secHighL} we review the first in our class of SDDM models. In this model, some of the DM consists of excited positronium-like bound states which are stable due to a large angular momentum barrier.  Interaction with the matter in the Earth can change the angular momentum of the bound state, rendering it unstable.  We then estimate the signal rate of this model in large neutrino detectors and discuss possibilities for its early Universe history. In the second model, presented in Section~\ref{sectunnelling}, the DM is in a metastable minimum far away from the origin of a bound-state potential. Interaction with the Earth can shake the internal structure and move the system to the global minimum where annihilation is very fast. In our last example, Section~\ref{secSymmetry}, DM consists of dark baryons.  A baryon number violating interaction with matter causes a transition from a dark baryon to a dark meson which can now decay in a detector. In Section~\ref{sec:pheno}, we discuss the phenomenological signals of our framework in large detectors further in order to motivate more detailed studies. We conclude and provide some further directions for exploration in Section~\ref{sec:conc}.

\section{High angular momentum stabilization}\label{secHighL}

In this model, a component of the DM is a positronium-like bound state which is stabilized by angular momentum conservation. In particular, the self-annihilation of this ``dark positronium'' is suppressed due to a large angular momentum barrier. An interaction with the Earth allows the dark matter to transition to a lower angular momentum state that has a lifetime which is exponentially shorter.

The dark sector contains light fermions $\chi$ and $\bar\chi$ which are charged under two broken abelian gauge groups $U(1)_\phi$ and $U(1)_V$. 
The Lagrangian for this sector takes the form
\beq
\mathcal{L} = \bar \chi i \slashed{D}\chi - m_\chi \bar\chi \chi 
- \frac{1}{4} \phi^{\mu\nu} \phi_{\mu\nu} + {1 \over 2} m_\phi^2 \phi_\mu\phi^\mu  
- \frac{1}{4} V^{\mu\nu} V_{\mu\nu} + \frac{1}{2} m_V^2 V_\mu V^\mu - \frac{\epsilon}{2} V_{\mu\nu} F^{\mu\nu} \ , 
\eeq
where 
\beq
D_\mu \chi = (\partial_\mu + i g_\phi \phi_\mu + i g_V V_\mu)  \chi \ , 
\eeq 
and $\phi_{\mu\nu}$, $V_{\mu\nu}$, and $F_{\mu\nu}$ are the field strengths for $\phi$, $V$ and the standard model photon respectively. For simplicity, we assume no kinetic mixing of $\phi$ to any of the other gauge bosons. We further define the fine-structure constants, $\alpha_V = g_V^2/(4\pi)$ and 
$\alpha_\phi = g_\phi^2/(4\pi)$, and the usual $\alpha_\mathrm{EM}=e^2/(4\pi)$.

The lighter of the two gauge bosons, $\phi$, is assumed to be light enough to mediate a long range force which leads to $\chi\bar\chi$ positronium-like bound states with $n_*\sim 10$ energy levels. On the other hand, $\phi$ is taken to be  too heavy to be emitted on-shell in transitions among energy levels. These constraints give
\begin{equation}\label{phimassrange}
\frac{1}{4}\alpha_\phi^2 m_\chi < m_\phi < \frac{1}{2n_*}\alpha_\phi m_\chi \, .
\end{equation}
We further assume that $V$ is too heavy to play a role in the bound state dynamics, but is light enough to be emitted in the self-annihilation of the $\chi\bar\chi$ state, thus
\beq
\frac{1}{2}\alpha_\phi m_\chi < m_V \lesssim m_\chi \ .
\eeq
We assume that $\alpha_V\gtrsim\alpha_\phi$ so that on-shell $V$s are produced in a sizable fraction of the annihilation processes. Our last assumption is
\beq\label{eq:mv>2me}
m_V > 2m_e \ ,
\eeq
so $V$ could decay into $e^+e^-$ and can be detected easily. 

Before proceeding to estimate lifetimes, we comment on some possible alternative models. For example, one could consider a model with a single vector boson from the dark sector playing the roles of both $\phi$ and $V$. However, in this case, Eq.~(\ref{eq:mv>2me}) with $\alpha_\phi \lesssim 10^{-2}$ and $n_{*} \sim 10$, and the upper bound in Eq.~(\ref{phimassrange}) can only be reconciled when $m_{\chi}\gg\,1\;\text{GeV}$. As we are interested in a wide range of $m_\chi$, as low as the MeV scale, see Section~\ref{BSscattering}, we keep both $\phi$ and $V$ in the following discussion. Another possibility is to allow $\phi$ to also mix with the SM photon, a possibility which, for the most part, affect the signal rate estimated in Section~\ref{BSscattering}. In addition, there may be other variants of this model where either or both $V$ and $\phi$ are replaced by light scalars or pseudoscalars. We focus here on the gauge case for concreteness but other options are considered when the phenomenology is discussed in Section~\ref{sec:pheno}.

\subsection{Bound state lifetimes}

We now consider the lifetimes of the $\chi\bar\chi$ bound states. As we show below, states with $\ell \gtrsim 10$ have lifetimes much longer than the cosmological bound on DM lifetime, while $S$-wave states annihilate very quickly, within the time it traverses the Earth.

In the vacuum, bound states could be destroyed either by direct $\chi\bar\chi$ annihilation or by de-excitation to a lower-$\ell$ state. We begin with an estimate of direct annihilation to $V$s. (In the following we also refer to $V$ as the dark photon.)
The number of $V$s in the final state is denoted by $N_V$. This number is 2 for bound states that are even under charge conjugation and 3 for odd ones. 
The leading decay to consider is thus $\Psi_{n,\ell} \to V V$ or $\Psi_{n,\ell} \to V V V$.
The amplitude for annihilation into dark photons for the $\Psi_{n,\ell}$ state is proportional to the $\ell$th derivative of the wave function at the origin. Roughly speaking every derivative result in an extra power of $\alpha_\phi$ in the amplitude.
Assuming no significant phase space suppression, these decay rates can be estimated by~\cite{An:2016gad},
\beq
\Gamma_{n,\ell \to V's} \sim \left({\alpha_\phi \over n} \right)^{2 \ell+3} \alpha_V^{N_V} m_\chi \ ,
\eeq
Annihilation into $\phi\phi$ or into $\phi V$ have a similar power counting. With
 $m_\chi \sim 1\,{\rm GeV}$, and $\alpha_V\sim\alpha_\phi\sim 10^{-2}$ we find that, for example, the inverse width of annihilation of the $\Psi_{n=10,\ell=9}$ state is about $10^{42}$ seconds, while $\Psi_{n=7,\ell=6}$ lives for about $10^{22}$ sec. The latter is too short to evade bounds on the lifetime of all of DM, but is safe if that state is a small fraction of the DM.

Next, we consider de-excitation. Since $m_V, m_\phi > \alpha_\phi^2 m_\chi/4$, the de-excitation of $\Psi_{n,\ell}$ to lower states cannot happen via on-shell $\phi$s or $V$s. Rather the leading decay is to three photons through an off shell $V$ or an off shell $\phi$, if it couples to the SM. The binding energy difference, of order $\sim\alpha_\phi^2 m_\chi$, is assumed to be too small to emit a pair of $e^+e^-$ for all the parameter space under consideration. There is also the possibility for the off-shell $V^*$ to turn into two neutrinos via its kinetic mixing with the SM $Z$-boson. The partial rate, in this case, is even more suppressed than the two or three photon final states we are considering.

The de-excitation rate of $\Psi_{n,\ell}\to \Psi_{n-1,\ell-1}$ via off-shell $V$ is approximately
\begin{eqnarray}
\Gamma_{n, \ell \rightarrow n-1,\ell-1} \sim \left( \frac{\alpha_\phi}{n^2} \right)^2 \frac{\alpha_V \epsilon^2 \alpha_\mathrm{EM}}{m_V^4} \frac{\alpha^3_{EM}}{m_e^8} 
\left( \frac{\alpha_\phi^2 m_\chi}{n^3} \right)^{13} \ ,
\end{eqnarray}
where the term in the first parentheses $\alpha_\phi/n^2$ is the matrix element for dark photon radiation from the bound states, and 
the term in the last parentheses $\alpha_\phi^2 m_\chi/n^3$ stands for the difference in binding energy. This dimensionful quantity is raised to the thirteenth power in order to make up the dimension of the rate.
For example, with $m_\chi \sim m_V \sim 1\,{\rm GeV}$, $\epsilon\sim10^{-2}$, and $\alpha_V \sim \alpha_\phi\sim 10^{-2}$ we find that the lifetime of the $\Psi_{n=10,\ell=9}$ state is about $10^{41}$ seconds, making it cosmologically stable. The lifetime is even longer for smaller values of $m_{\chi}$, or for transitions with $\Delta \ell > 1$.

If $\phi$ also mixes with the photon, the bound state de-excitation rate via off-shell $\phi$ can be estimated in a similar way, with the $V$ kinetic mixing parameter and the dark photon mass replaced by those of $\phi$. If $\phi$ is a scalar particle and does not mix with the SM Higgs boson, the de-excitation is even slower because it requires at least three loops and two powers of the kinetic mixing parameter $\epsilon$ for it to decay into two SM photons.

We have thus shown that high-$\ell$ excitations of the $\chi\bar\chi$ bound state are cosmologically stable and are a viable candidate to be at least a portion of DM.  
On the other hand, the bound states with small $n$ and $\ell$ quantum numbers can decay much faster, into $2V$ or $3V$, depending on its charge-conjugation parity ($C=(-1)^{L+S}$) as discussed above. The amplitude for an $S$-wave $\chi\bar\chi$ bound state to annihilate into $2V$ is proportional to its wave function at the origin, and, neglecting phase space effects, the decay rate is
\begin{eqnarray}
\begin{split}
&\Gamma_{\Psi^{(+)}_{n,0}\to 2V} =\frac{\alpha_\phi^3 \alpha_V^2 m_\chi}{2n^3} \, , \\
&\Gamma_{\Psi^{(-)}_{n,0}\to 3V} = \frac{2(\pi^2-9)\alpha_\phi^3 \alpha_V^3 m_\chi}{9\pi n^3} \, .
\end{split}
\end{eqnarray}  
Setting $m_\chi=1\,$GeV, $\alpha_V=\alpha_\phi=0.01$, and a DM speed of $v \sim 10^{-3}$ we find the decay length of the ground state to be, $v\tau_{\Psi^{(+)}_{1,0}\to 2V} \sim 4\times 10^{-7}$\,cm, or $v\tau_{\Psi^{(-)}_{1,0}\to 3V} \sim 3\times 10^{-4}$\,cm, which is a prompt decay. For an S-wave state with $n\leq10$ the decay is prompt in view of DM or neutrino detectors.

In conclusion, the model has the two desirable ingredients: a cosmologically long lived state and a short lived one that can generate a signal in the detector. The next task is to estimate the rate for transitioning from the long to the short lived states.

\subsection{Bound state scattering}
\label{BSscattering}

In this subsection, we demonstrate how collisions of a high-$\ell$, long-lived bound state $\Psi$ with a nucleus can induce a transition to a lower $\ell$ state $\Psi'$ that decays quickly into SM particles.
The amplitude for the leading non-relativistic inelastic scattering of a proton with a bound state  
\beq
\Psi(\vec{p}_1)+ p(\vec{p}_2) \to \Psi'(\vec{p}_3)+ p(\vec{p}_4),
\eeq
occurs via $t$-channel dark photon ($V$) exchange and it
takes the form
\begin{eqnarray}
\mathcal{M}_{\Psi + p\to \Psi' + p} = 4 m_\chi F_D(|\vec{q}\,|) \frac{g_V \epsilon e}{|\vec{q}|^{\,2} + m_{V}^2} \bar u_p(\vec{p}_4) \gamma^0 u_p(\vec{p}_2) \ .
\end{eqnarray}
Here $u_p$ is the proton spinor, and $F_D(|\vec{q}\,|)$ is the form factor for the $\Psi\to\Psi'$ transition in the presence of a three-momentum $\vec{q} = \vec{p}_3-\vec{p}_1$ injection,
\begin{eqnarray}\label{eq:ff}
F_D(|\vec{q}\,|) = \int d^3 {\vec x} \,  \Psi'^*(\vec{x}) \Psi(\vec{x}) \left[ e^{{i \vec{q}\cdot \vec{x}}/{2}} - e^{{-i \vec{q}\cdot \vec{x}}/{2}} \rule{0mm}{4mm}\right] \ .
\end{eqnarray}
In the following we estimate $F_D(|\vec{q}\,|)$ by making the approximation that $\Psi$ and  $\Psi'$ are Coulomb-like wave functions. 
Parity invariance implies that the orbital angular quantum numbers of the initial and final states in Eq.~(\ref{eq:ff}) must differ by an odd integer.

The differential scattering cross section off a nucleus (with atomic and mass numbers $Z, A$) takes the approximate form
\begin{eqnarray}\label{eq:cs}
\frac{d \sigma_{\rm scatter}}{d|\vec{q}\,|^2} \simeq \frac{g_V^2 \epsilon^2 e^2}{4 \pi v^2(|\vec{q}|^{\,2} + m_{V}^2)^2} \times \left|F_D(|\vec{q}\,|)\right|^2 \times Z^2 F^2(|\vec{q}|) \ ,
\end{eqnarray}
where $v\simeq 10^{-3}$ is the incoming DM bound state velocity. For $F(|\vec{q}|)$,  the nuclear form factor, we use the Woods-Saxon form given in~\cite{Jungman:1995df}.
Note that when $\Psi'$ is a deeper bound state than $\Psi$, both the binding energy difference and the initial kinetic energy contribute to
the momentum transfer, $\vec q$. In the small $v$ limit, the upper and lower limit of $|\vec{q}|^2$ take the form
\beq
\left(|\vec{q}|^2 \right)_{\rm min}^{\rm max} =2 \mu_{\Psi' A} (m_\Psi - m_{\Psi'}) \pm \mu_{\Psi' A} v \left( \frac{m_\Psi}{m_{\Psi'}}\right) \sqrt{8 \mu_{\Psi' A} (m_\Psi - m_{\Psi'})} + \mathcal{O}(v^2) \ ,
\eeq 
where the reduced mass $\mu_{\Psi' A}$ is defined as $\mu_{\Psi' A} = m_{\Psi'} m_A/(m_{\Psi'} + m_A)$.
For weakly bounded states, we have $m_\Psi \sim m_{\Psi'} \simeq 2 m_\chi$ and $m_\Psi - m_{\Psi'} \sim m_\chi \alpha_\phi^2$.
The above expansion converges for $\alpha_\phi\gtrsim v$, {\it i.e.}, when the binding energy release, $m_\Psi - m_{\Psi'} \sim m_\chi \alpha_\phi^2$, dominates over the DM kinetic energy, $\sim m_\chi v^2$. 
To further simplify the result, we consider the limit $|\vec{q}^2|\ll m_V^2$ and $m_\Psi\ll m_A$, which is the region of interest for this study. In this case, the scattering cross section is approximately,
\begin{eqnarray}\label{eq:cscompact}
\sigma_{\rm scatter} \simeq \frac{64\pi \epsilon^2 \alpha_\mathrm{EM} \alpha_V \alpha_\phi m_\chi^{2}}{\pi v m_{V}^4}\times \left|F_D(|\vec{q}\,|)\right|^2 \times Z^2 F^2(|\vec{q}|) \ ,
\end{eqnarray}
with $F_D(|\vec{q}\,|)$ and $F(|\vec{q}\,|)$ evaluated at the typical momentum transfer, $|\vec{q}| \sim m_\chi \alpha_\phi$.
Numerically, $F(|\vec{q}|)\simeq 1$ and the typical value of $F_D(|\vec{q}\,|)$ is around $10^{-2}$ for $\alpha_\phi=10^{-3}-10^{-2}$. 

Before proceeding to estimate a signal rate, we note that if the lighter vector boson $\phi$ were allowed to have a kinetic mixing $\epsilon_\phi \phi_{\mu\nu}F^{\mu\nu}$, the scattering rate in Equation~(\ref{eq:cscompact}) would have an additional contribution with $m_V$ and$\epsilon$ replaced by $m_\phi$ and$\epsilon_\phi$. This contribution can potentially be larger because $\phi$ is lighter. In this sense the rate estimate in the next subsection, where $\epsilon_\phi$ is set to zero, is conservative.

\subsection{Signal Rates in Neutrino Detectors}\label{sec:signal-rate}

Once a high-$\ell$ state hits the Earth, it scatters and transitions to a low-$\ell$ state which is short-lived. Generically, the short-lived state decays promptly to two or three $V$s, both of which can have signals in a large detector. The signal event rate depends primarily on the rate of transitions from long- to short-lived states. It takes the familiar form 
\beq
\Gamma_\mathrm{signal} = N_{T}^{(\mathrm{eff})} n_{\Psi} \left<\sigma v \right>_{\text{scatter}} \ ,
\eeq
where $N_T^{(\mathrm{eff})}$ is the effective number of target nuclei which is defined below and $\left<\sigma v \right>_{\text{scatter}}$ is the velocity average of the cross section to scatter and transition from the high to the low-$\ell$ state.

The precise nature of the signal and the effective number of target nuclei depend on the lifetime of the short lived state $\Psi'$ and and the mediator $V$. It is convenient to define the average decay lengths for these two states
\begin{eqnarray}
L_{\Psi'} = \gamma_{\Psi'} v_{\Psi'} \tau_{\Psi'} \ , \qquad
L_{V} = \gamma_V v_V \tau_V \ , \label{eq:LV}
\end{eqnarray}
where the $\tau$s are the respective lifetimes of the two particles,  $\gamma_{\Psi'}\approx 1$ and $v_{\Psi'}\sim 10^{-3}$ are
 the typical boost and velocity of the meta-stable state, while $\gamma_V$ and $v_V$ are the boost and velocity of $V$ which are set in any given kinematics. In the case of $\Psi'\to 2V$ decay, 
\beq
L_V= c\tau_V \sqrt{{m_{\Psi'}^2}/{4 m_V^2} -1}.
\eeq

We can now estimate the effective number of target nuclei $N_T^{(\mathrm{eff})}$. If both decay lengths are  shorter than the typical size of the detector, $L_{\Psi'}, L_V < L_\mathrm{det}$ the calculation is simple. Since neither particle can travel far the transition scattering has most likely happened inside the detector and the number of targets is simply the number of nuclei in the detector
\begin{eqnarray}\label{eq:rate-short-lived}
L_{\Psi'}, L_V < L_\mathrm{det}\,:\qquad\qquad  
N_T^{(\mathrm{eff})} \sim n_\mathrm{det}\,V_\mathrm{det} \ ,
\end{eqnarray}
where $n_\mathrm{det}$ and $V_\mathrm{det}$ are the number density of nuclei and the volume of the detector. In this case, the coherence factor in scattering, the $Z^2$ factor in Eq.~(\ref{eq:cscompact}), should be taken as that of the detector material, be it water, liquid argonne, etc.


In the case that either of the decay lengths is larger than the typical detector size, $L_{\Psi'} >L_\mathrm{det}$ or $L_V>L_\mathrm{det}$, the production of $V$s is likely to occur elsewhere in the Earth. It is interesting that even in this case, the effective number of target nuclei is parametrically the same as in the short lifetime case.
Consider, for example, the case where $L_V$ is much longer than the detector but shorter than the size of the earth, while $L_{\Psi'}$ is short. In this case, we can assume the decay of $\Psi'$ to $V$s occurred at the scatter site.
The effective number of targets is a sum over Earth nuclei, weighted by the probability that a $V$ decays inside the detector given that is originated near the nucleus in question
\begin{eqnarray}\label{eq:rate-long-lived}
\begin{split}
L_\mathrm{det} < L_V \lesssim R_{\oplus} \,:\qquad  
N_T^{(\mathrm{eff})} &\sim  
n_\oplus \int_{\oplus\cap L_V} d^3\vec r_T 
\frac{L^2_\mathrm{det}}{4\pi |\vec r_\mathrm{det}-\vec r_T|^2 }\frac{L_\mathrm{det}}{L_V} 
\sim n_\oplus\,V_\mathrm{det} \ ,
\end{split}
\end{eqnarray}
where the integration limits are 
\beq
|\vec r_T| < R_\oplus  \quad\mathrm{and}\qquad 
 0< |\vec r_\mathrm{det}-\vec r_T| <L_V \eeq
in a coordinate system in which the origin is at the center of the Earth.
Here we have treated the earth as a sphere of constant density $n_\oplus$. 
The fraction in the integrand is the suppression due to the small solid angle that the detector subtends, as seen from the scatter site. In the last step, we have used the fact this suppression factor is undone by the fact that the number of target nuclei grows like $|\vec r_\mathrm{det}-\vec r_T|^2$ (which will show up as a Jacobian). This effect is similar to the signal rate in~\cite{Feldstein:2010su}. Note that here we do not account for an $O(1)$ geometrical factor, which, for example, comes from the fact that there is more Earth  below the detector than above it.
The case in which both $L_V$ and $L_{\Psi'}$ are larger than the detector is slightly more convolved but parametrically similar.
Note, that in cases of long lived $V$ and/or $\Psi'$ the coherence factor of $Z^2$ would be the averaged over the rock, perhaps dominated by iron, but may get contributions from heavier elements such as lead. In our simplified analysis we took Earth's composition to be $100\%$ water, which is valid up to an $\mathcal{O}(1)$ factor.

Though the signal rate does not depend strongly on $L_V$, the character of the signal does. In the case of a short lived $V$, $L_V<L_\mathrm{det}$, both $V$s can decay inside the detector and will result in \emph{two} back-to-back or \emph{three} $e^+e^-$ (or $\mu^+\mu^-$, if kinematically allowed) pairs. In the case where  $L_V > L_\mathrm{det}$ it is more likely that just one such pair will be in the detector. In the case where $L_V$ is much longer than the underground depth of the laboratory, there will be more events coming from below than from above. 
These, and other signal characteristics are discussed in more detail in Section~\ref{sec:pheno}.

\begin{figure}[t]
\centerline{\includegraphics[width=0.9\textwidth]{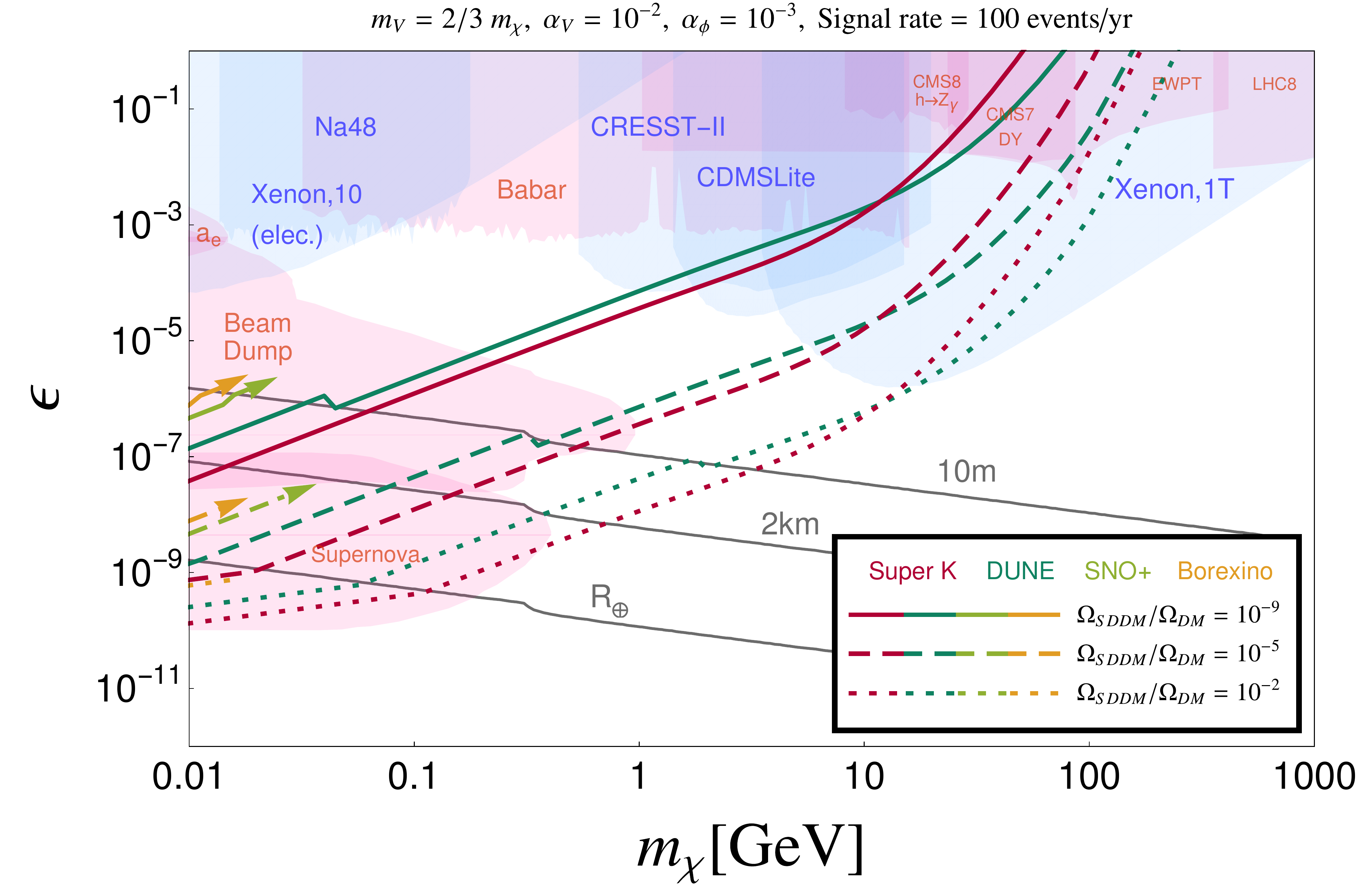}}
\caption{ Contours that give 100 DM self-destruction events per year in the $\epsilon-m_\chi$ parameter space of the angular momentum model. The other parameters are chosen to be $m_V=2m_\chi/3$, $\alpha_V=0.01$, $\alpha_\phi=0.001$. The colorful thick curves correspond to Super-Kamiokande (red), and DUNE (dark green). Contours for SNO+ (light green), Borexino (orange) are shown as arrows in order to reduce the number of curves with the understanding that they run parallel to that of Super-K. The solid, dashed and dotted curves assume the initial DM bound state is $\Psi_{10,9}$ and comprises $10^{-9}$, $10^{-5}$ and $10^{-2}$ of the total DM relic density, respectively. The gray curves correspond to constant decay length contours of the dark photon $V$ (see Section~\ref{sec:pheno} for their effect on the signal characteristics). 
The light red and blue shaded regions show the existing experimental constraints from searching for visibly-decaying dark photons and DM direct detection (assuming $\chi$ to be the dominant DM), respectively.}
\label{moneyplot}
\end{figure}

The reach of a particular experiment depends on the detector-specific background rate for the signal in question. Though we expect very little background that mimics our particular signals, we did not delve into these detector-specific details. Instead, we simply estimate the event rates in several detectors as a function of the model parameters. 
In Fig.~\ref{moneyplot}, we show the parameters that lead to 100 events per year in the $\epsilon - m_\chi$ plane. 
For a weakly-coupled bound state as the SDDM, its mass is related to that of $\chi$, $m_{\Psi} \simeq 2 m_\chi$.
The colorful thick curves correspond to the observation of this rate of $\Psi'$ decay events in various neutrino experiments: Super-Kamiokande, SNO+, Borexino and DUNE. In order to reduce the clutter in the plot, we have denoted the limits on the smaller two experiments with arrows, with the understanding that their rate contours run parallel to those of Super-K.
The 100-event-per-year contour for $\epsilon$ is given by:
\beq
\epsilon_{100}^2=\frac{100\mbox{ events}}{T_{\text{year}} \times n V
\times n_{\chi\bar\chi}
\left<\sigma v \right>^{(\epsilon=1)}_{\text{scatter}}\times \text{Br}\left(V\rightarrow l^+l^-\right)}\, ,
\eeq 
where $nV$ is the effective number of targets, defined by the terms in parentheses in Eqs.~(\ref{eq:rate-short-lived}) and (\ref{eq:rate-long-lived})  for $L_{V}<L_\mathrm{det}$ and $L_{V}>L_\mathrm{det}$ respectively. Our result is a rough estimate as we have omitted the $\mathcal{O}(1)$ geometry-dependent factor for $L_V>L_\mathrm{det}$ and we neglected the density variations in the Earth.
The curves in Fig.~\ref{moneyplot} are derived assuming that the initial DM bound state is $\Psi_{10,9}$ and comprises $10^{-9}$, $10^{-5}$ or $10^{-2}$ of the total DM relic density. For the final state $\Psi'$, we consider all possible $S$-wave bound states with $n=1, \dots,9$. (Note that this is a conservative assumption as we did not consider the effect of any non-$S$-wave states.) The model parameters in the plot are set to: $m_V=(2/3)\, m_\chi$, $\alpha_V=0.01$ and $\alpha_\phi=0.001$.
Since we have used $\alpha_{\phi} \ll \alpha_{V}$ the resulting unstable bound state annihilates predominantly into two $V$s (and then into SM particles). 
In the shaded regions, we also show the existing experimental constraints from searching for visibly-decaying dark photons~\cite{Alexander:2016aln} as well as direct detection (assuming $\chi$ to be the dominant DM). Interestingly, the signals can help to probe a large portion of the parameter space that is not yet accessible otherwise with the existing dark photon searches. This gives a strong motivation for exploring the neutrino experimental data and looking for the existence of DM bound states in nature.
Although the current neutrino detectors focus on energy ranges relevant for incoming neutrinos~\cite{Richard:2015aua, Andringa:2015tza, Bellini:2008mr, Acciarri:2015uup}, they are also sensitive to higher energy ranges and might be used to probe SDDM up to higher masses. We will return to a more detailed discussion of the phenomenology of SDDM and some search strategies in Section~\ref{sec:pheno}.

\subsection{Bound state production in the early universe}

To complete our picture of angular momentum stabilized dark matter, we discuss some possibilities for the production of high-$\ell$ bound states in the early universe. As we will see, the cosmology in this model is challenging and might involve a non-thermal production or other mechanisms.
The difficulty arises from the constraint Eq.~(\ref{phimassrange}), which implies that the mediator $\phi$ is heavier than the Rydberg energy, $m_\phi>(\alpha_\phi^2 m_\chi)/4$. This implies that even at zero temperature, an incoming mediator has the potential to dissociate the $\chi\bar{\chi}$ bound states. This fact strongly restricts the production mechanisms of $\chi, \bar\chi$ bound states in the early universe. 
We now discuss several possible production mechanisms. We leave a more dedicated study of these possibilities for a future work.

{\it Thermal freeze out}. 
Consider the case where $\chi$, $\bar\chi$,  and $\phi$ are in thermal equilibrium with each other in the early universe. 
For simplicity, we assume that the dark sector has the same temperature as the SM sector.
Around $T\sim m_\chi/25$, the $\chi$ and $\bar\chi$ particles freeze out by annihilating into $\phi$s and/or $V$s. Assuming $\alpha_\phi\sim\alpha_V$, their relic abundances satisfy
\beq\label{omegachi}
\Omega_\chi = \Omega_{\bar \chi} \sim 10^{-4} \left( \frac{\alpha_\phi}{0.01} \right)^{-2} \left( \frac{m_\chi}{1\,\rm GeV} \right) , \qquad
\frac{n_\phi}{n_\chi} \sim 10^{14} \left( \frac{\alpha_\phi}{0.01} \right)^{2} \left( \frac{m_\chi}{1\,\rm GeV} \right)^{-1} \ .
\eeq
After freeze out, $\chi,\bar\chi$ bound states could be formed through the process $\chi+\bar\chi \to \left(\chi\bar{\chi}\right) + \phi$, where $\left(\chi\bar{\chi}\right) $ is a $\chi\bar{\chi}$ bound state. 
Since the formation rate is smaller than the Hubble expansion, the resulting bound state relic abundance is given by
\beq\label{omegabs}
\Omega_{BS} = \Omega_\chi \frac{R_{BSF}}{H}, \qquad R_{BSF} = n_\chi \sigma_{BSF} \ ,
\eeq
where $\sigma_{BSF}$ is the bound state formation cross section.
On the other hand, a bound state could be dissociated by encountering a $\phi$ particle, $\left(\chi\bar{\chi}\right) + \phi\to \chi +\bar\chi$. The dissociation cross section $\sigma_{DIS}$ is related to $\sigma_{BSF}$ via crossing symmetry. For $T\gtrsim m_\phi$, the two are related by $\sigma_{DIS}\sim (m_\chi/T)^{3/2} \sigma_{BSF}$~\cite{Wise:2014jva}. 
Note that because the $\phi$ mass is greater than the binding energies, see Eq.~(\ref{phimassrange}), we must rely on the kinetic energies of $\chi, \bar\chi$, dictated by the temperature $T$, for bound state formation to occur.
 The dissociation rate for each bound state is then
\beq
R_{DIS} = n_\phi \sigma v_{DIS} \sim \left( \frac{n_\phi}{n_\chi} \right) \left( \frac{m_\chi}{T} \right)^{\frac{3}{2}} R_{BSF} \ .
\eeq
From Eqs.~(\ref{omegachi}) and (\ref{omegabs}), we learn that
the probability for a bound state to dissociate satisfies,
\beq
P_{DIS} = \frac{R_{DIS}}{H} \gtrsim 10^8 \left( \frac{m_\chi}{T} \right)^{\frac{3}{2}} \left( \frac{\alpha_\phi}{0.01} \right)^{4} \left( \frac{m_\chi}{1\,\rm GeV} \right)^{-2} \ .
\eeq
The above result implies that all bound states are quickly destroyed within a Hubble time as soon as they are produced.

This is a problem since we would like
$\Omega_{BS} > 10^{-10}$ for the SDDM to be of experimental interest (see Fig.~\ref{moneyplot}). This problem
could be cured if we further extend the dark sector with light states which $\phi$ can annihilate or decay into. The coupling of these light states to $\phi$ has to be large enough to deplete the $\phi$ population and prevent bound-state dissociation. Its coupling to the $\chi$, has to be small enough to not dissociate the bound states themselves. 

{\it $\chi$ as asymmetric DM}. Since $\chi$ is much lighter than the weak scale, the thermal production scenario cannot accommodate a situation where $\chi$ comprises all the DM relic abundance. It is then tempting to consider the production of $\chi$ in an asymmetric scenario. Additionally, in such a scenario one can avoid cosmological constraints on DM annihilation such as the CMB~\cite{Madhavacheril:2013cna}. The problem with this scenario is that
the number of $\bar\chi$ is exponentially suppressed at late time~\cite{Graesser:2011wi, Gelmini:2013awa} because the $\chi\bar\chi$ annihilation cross section is very large.
The exponential suppression of $\bar\chi$ would make it even harder to form any $\chi\bar\chi$ bound states in the early universe.

{\it Non-thermal history}. The thermal freeze out picture we discussed above does not work in its minimal form mainly because it produces many more $\phi$ particles than $\chi$ and $\bar\chi$.
This problem could be alleviated if the dark sector particles are produced non-thermally, with a suppressed $n_\phi/n_\chi$ ratio compared to Eq.~(\ref{omegachi}). Non-thermal production mechanisms include the freeze-in of $\chi$ and $\bar\chi$ through a small kinetic mixing between $V$ and/or $\phi$ and the SM photon, or the late decay of some heavier dark sector states.
The relic abundance of high-$\ell$ bound states in these cases is model dependent but it seems plausible that they can still be probed in neutrino detectors.

{\it Additional confining dark force}. Another way of efficiently producing the high-$\ell$ bound states is to replace $\phi$ by a confining dark force. This implies that $\chi$ is colored under a non-abelian dark gauge interaction. In this case, there are no free $\chi$ and $\bar\chi$ particles at distances larger than the inverse of the dark confinement scale. In that case, it is likely for the high-$\ell$ $\chi\bar\chi$ states to form efficiently~\cite{Kang:2006yd}. If the dark confinement scale of this gauge interaction is very small, the low-level states are to a good approximation given by a $1/r$ potential. Then, the self-destructing DM scattering, which involves short-distance physics, and the experimental detection discussions in section~\ref{BSscattering} still remain valid. 

{\it Late production of bound states}. In addition to the production in the early universe, we can consider the more recent production of $\chi\bar\chi$ bound states by a late time binding process of relic $\chi$ and $\bar\chi$. 
The $\phi$ particles are sufficiently diluted in the universe today so they no longer play a role in dissociation. 
Because $m_\phi$ is larger than the Rydberg energy, the $\chi$ and $\bar\chi$ in the initial state must have large enough velocities.
One might imagine more efficient bound state formation inside neutron stars where $\chi$ and $\bar\chi$ are gravitationally accelerated.
The formed bound states could be liberated if the neutron star belongs to a binary system which inspirals and finally merges~\cite{TheLIGOScientific:2017qsa}. 
This picture then ties the bound state formation rate to the neutron star merger rate.

\section{Tunneling stabilization}\label{sectunnelling}

We now consider a different mechanism to generate the meta-stable state of DM required for the self-destructing phenomenology. The idea is that the long-lived DM state is protected by a potential barrier which keeps the constituents apart, requiring exponentially suppressed tunneling to self annihilate. The same bound state can scatter into a short-lived bound state, where the constituents are localized closer together. This transition can happen through excitation to a de-localized state and a decay to the localized state near the origin. 

A key ingredient in this tunneling stabilization mechanism is the existence of a metastable minimum far from the origin, so that the tunneling probability is exponentially small. This happens in nature usually due to an interplay between several forces, both attractive and repulsive. Consider for example the states that can be made of two deuterium atoms. The two atoms can be assembled into a di-deuterium molecule, $D_2$. The two nuclei of $D_2$ can, with a very long lifetime, tunnel through the Coulomb barrier to form a Helium-4 atom. Similarly, some common isotopes of table salt, NaCl, can tunnel to become a Nickel atom since they have the same number of protons, neutrons, and electrons. Though these examples are inspiring, they do not quite fulfill our requirements since excited molecular states of $D_2$ are still very unlikely to fuse into helium. Perhaps real-world examples which are closer to what self-destructing DM requires are super cooled systems in crystals and liquids~\cite{supercool}.~\footnote{For further inspiration, click here 
$\href{https://www.youtube.com/watch?v=fXZrzMPdBNo}{\star}$.}

\begin{figure}
\centering
\includegraphics[height=6cm]{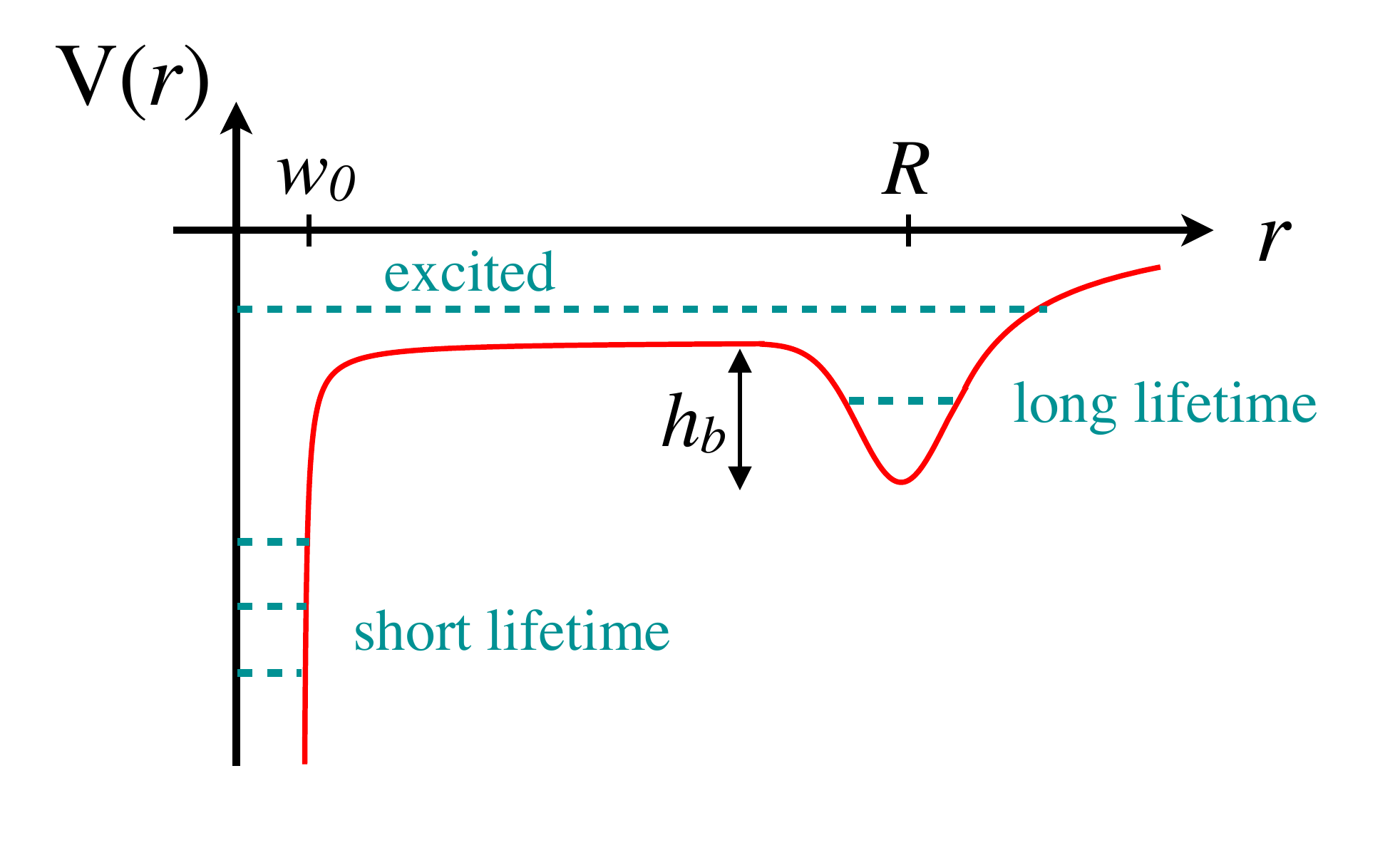}
\caption{An illustration of a two minima potential as described in Section~\ref{sectunnelling}.}
\label{fig:sketch}
\end{figure}

To illustrate the main features of tunneling stabilization, we consider a different kind of $\chi\bar{\chi}$ bound state, one which is bounded not by a Coulomb mediator but by a potential $V(r)$ such that $r$ is the relative distance between the $\chi$ and $\bar\chi$. We remain agnostic as to the origin of this potential, but one can imagine an effective potential due to several mediators. Furthermore, we consider the case in which $V(r)$ has a global minimum at $r=0$ and a local minimum at $r=R$, with the two separated by a potential barrier as shown in Fig.~\ref{fig:sketch}. 
This potential is reminiscent of a molecular potential, but with the coulomb barrier truncated by hand. 
We call a potential of this sort a two minima potential.
This potential has three kinds of states: 
\begin{itemize}
\item Long lifetime states centered around the local minimum at $r=R$.
\item Short lifetime states centered around the global minimum at $r=0$.
\item Excited states above the barrier which are de-localized within the full potential well.
\end{itemize}
Just like the angular momentum case, the $S$-wave decay rate of a $\chi\bar{\chi}$ state is proportional to its wavefunction squared at the origin
 \begin{eqnarray}\label{eq:gammapos}
\Gamma_{\chi\bar{\chi}}=4\pi\,\frac{\alpha^2_V}{m^2_\chi}\,{\left|\psi_{\chi\bar{\chi}}(0)\right|}^2\, .
\end{eqnarray}
The states localized at $r=0$ have significant support at the origin and so their lifetime is very short. In contrast to these, the states localized at $r=R$ have an exponentially small support at the origin and so their lifetime is very long. Additionally, the overlap of the long and short lifetime eigenfunctions is negligible, and so is the rate of spontaneous emission from one to the other.
The excited states are unsuppressed at both $r=0$ and $r=R$ so they quickly either annihilate or spontaneously emit a dark photon and relax to a lower state. When the depth of the global minimum is much larger than the local minimum, the number of short lifetime states is much larger than the number of long lifetime states ($N_{\text{Short}}\gg N_{\text{Long}}$). In that limit, the excited state decays predominantly to short lifetime states, which annihilate instantaneously.

For our SDDM scenario, we envision a fraction of DM is made up of the long-lived state with $\langle r\rangle \sim R$. An interaction of this state with a nucleus on Earth can excite it to the delocalized state, which then proceeds to decay and annihilate as discussed above. The annihilation products, assumed to be dark photons, can then decay in a detector leading to a signal. In the Appendix, we show a toy example which we used to estimate the various rates more quantitatively. We find that the lifetimes of the long, short and excited states scale as
\begin{eqnarray}
\frac{\tau_{\text{Short}}}{\tau_{\text{Long}}} \sim \frac{\tau_{\text{Excited}}}{\tau_{\text{Long}}}  \sim e^{-2\sqrt{m_\chi h_b}R} \, ,
\end{eqnarray}
where $h_b$ is the height of the potential barrier. 

In this paper, we do not construct a full model which yields a potential similar to the one in Fig.~\ref{fig:sketch}. A working model, however, would need to have both a significant barrier, say with $\sqrt{h_b}R\sim 20$, and excited bound states that can live classically above the barrier. Molecules found in nature have the former, but not the latter. 

While we were unable to construct a full model, we do not see any fundamental reason that we will not be able to. We can speculate about ways to modify the molecular picture to make it viable to our model. One way is to add a confining potential that will make the probability to decay into the short lived state significantly. Another idea is to have
droplets of a super-cooled dark fluid which make up a part of DM.

\section{Symmetry stabilization}\label{secSymmetry}

As a third class of models, we consider bound states in an asymmetric DM scenario, with the DM charged under baryon number. 
The self-destructing DM picture can be realized when the bound state interacts with SM nuclei.

\begin{figure}[t]
\includegraphics[height=3.5cm]{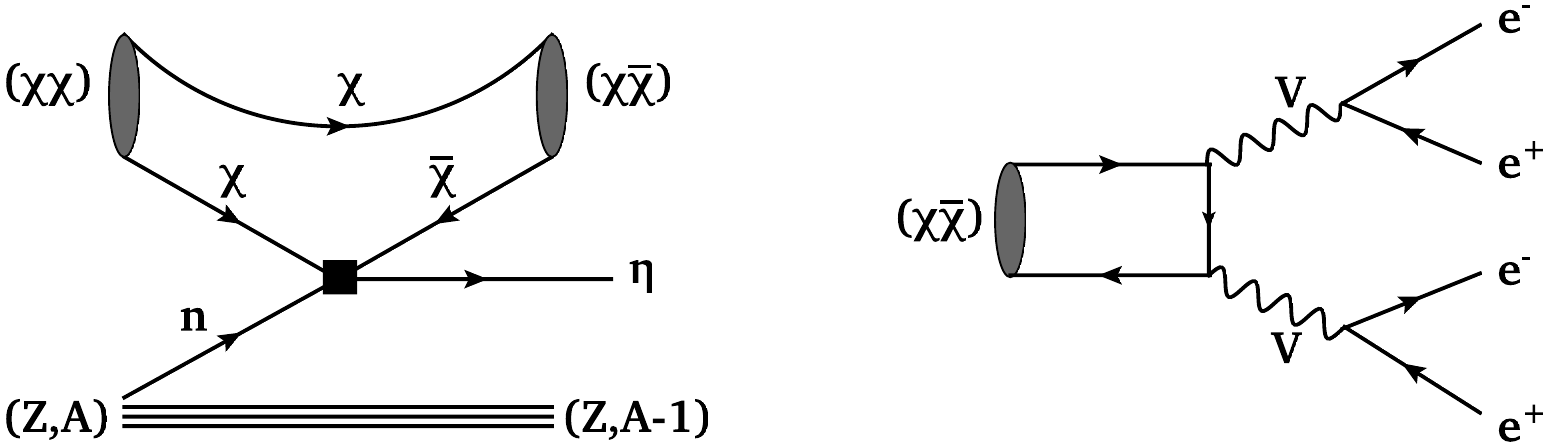}
\caption{{\it Left}: Feynman diagram for the scattering process $(\chi\chi) + n \to (\chi\bar\chi) + \eta$ in the model described in Section~\ref{secSymmetry}.
{\it Right}: A possible way of destructing the $(\chi\bar\chi)$ bound state after the scattering.}
\label{diagram_symmetry}
\end{figure}

More specifically, consider a Dirac fermion $\chi$ which carries the baryon number $q$, with $q\neq \pm1, \pm1/2$. The exchange of a scalar dark force carrier $\phi$ (with zero baryon number) allows both $\chi,\chi$ and $\chi,\bar\chi$ to form bound states. Because the $(\chi\chi)$ bound state is stable, we can assume it
comprises a significant fraction or even makes up all of the total DM relic abundance today.\footnote{With a scalar dark force, it is possible for a large number of $\chi$ particles to form stable bound states (also called nuggets)~\cite{Wise:2014jva, Wise:2014ola, Gresham:2017zqi}. In principle, a similar phenomenon as discussed below could also apply to nuggets. For simplicity, we only consider two-body bound states here.} We also introduce another fermion $\eta$ which carries non-zero baryon number $1+2q$. 
The assigned baryon quantum numbers above allow a dimension 6 effective interaction between $\chi, \eta$ and the neutron $n$,
\begin{eqnarray}
\mathcal{L}_{int} = \frac{(\overline{\chi^c} \chi) (\bar \eta n)}{\Lambda^2} + {\rm h.c.} \ .
\end{eqnarray}
If $2m_\chi<m_\eta+m_p$, the $(\chi\chi)$ bound state is stable. However, when it reaches the detector, the above interaction triggers the following reaction,
\begin{eqnarray}
(\chi\chi) + n \to (\chi\bar\chi) +\eta \ ,
\end{eqnarray}
as shown by the Feynman diagram in Fig.~\ref{diagram_symmetry} (left). 
At the ``partonic level'' in the dark sector, one of the $\chi$s participates in the reaction $\chi+n\to \bar \chi + \eta$. The other $\chi$ is a spectator, and it becomes bounded with the $\bar\chi$ in the final state into a $(\chi\bar\chi)$ bound state.
After this reaction, a neutron is removed from the target and the final state particle $\eta$ can escape away without further interaction. We consider $m_\eta \approx m_n$, and then the cross section for the scattering of a $(\chi\chi)$ ground state into a $(\chi\bar\chi)$ ground state is,
\begin{eqnarray}
\sigma_{(\chi\chi) + n \to (\chi\bar\chi) + \eta} \simeq \frac{\mu_{\chi n}^2}{\pi\Lambda^4} \ ,
\end{eqnarray}
where $\mu_{\chi n}=m_\chi m_n/(m_\chi+m_n)$, and we make the approximation that the dark form factor, in analogy to Eq.~(\ref{eq:ff}), is of order 1, when both $(\chi\chi)$ and $(\chi\bar\chi)$ are in the ground states.

Like the other two scenarios we discuss before, the most striking signal is generated due to the destruction of the unstable $(\chi\bar\chi)$ state. It can self-annihilate into SM particles via mediators ({\it e.g.}, a pair of dark photons, see Fig.~\ref{diagram_symmetry} (right)) between the two sectors. The present data from neutrino and DM experiments can be interpreted as constraints on the cutoff scale $\Lambda$. As an estimate, if the $(\chi\chi)$ bound state comprises all the DM relic abundance, having 100 events per year from $(\chi\chi)\to (\chi\bar\chi)$ scattering and the $(\chi\bar\chi)$ destruction inside the Super-K or DUNE detectors corresponds to a cutoff scale
\begin{eqnarray}
\Lambda \simeq 50 \,{\rm TeV} \left( \frac{m_\chi}{1\,{\rm GeV}} \right)^{1/4} \ .
\end{eqnarray}

An additional signature predicted in this model besides DM self destruction is the removal of a neutron from a target nucleus.
The resulting nucleus is often unstable which leads to additional hadronic activities inside the detectors.

As a necessary condition for this scenario to work, the $\eta$ particle mass must lie in the following window,
\begin{eqnarray}
m_n - 2m_\chi + BE_{(\chi\chi)} < m_\eta < m_n + KE + BE_{(\chi\chi)} - BE_{(\chi\bar\chi)} \ ,
\end{eqnarray}
where $m_n$ is the effective neutron mass inside the nucleus, $BE_{(\chi\chi), (\chi\bar\chi)}$ is the binding energy for the $(\chi\chi), (\chi\bar\chi)$ bound states, respectively,
and $KE$ is the kinetic energy of the incoming $(\chi\chi)$ bound state that enters the detector.
The upper bound of the above window is derived for the process $(\chi\chi) + n \to (\chi\bar\chi) + \eta$ to be kinematically allowed, and the lower bound is from the requirement of neutron stability against the decay $n\to \eta + (\bar \chi \bar\chi)$.
In addition, if $m_\eta \ll m_n$, the final state $\chi, \bar\chi$ particles would carry too large a relative momentum for them to remain bounded together. This limitation could be relaxed if the dark force for binding the $(\chi\bar\chi)$ and $(\chi\chi)$ states is not from the exchange of a scalar $\phi$ but a confining non-abelian ($SU(2)$) dark gauge interaction. In this case, there are no free $\chi, \bar\chi$ particles and bound states of them always form.

The idea of DM eliminating a neutron and rendering a nucleon unstable has been discussed previously in~\cite{Davoudiasl:2010am, Davoudiasl:2011fj, Huang:2013xfa}. However, in those models, the energy release is completely due to the destruction of a nucleon, yielding a signal very similar to the usual nucleon decay.
In contrast, in the model discussed here, as well as the two SDDM models presented in the previous sections,
the visible energy is released from $\chi \bar\chi$ annihilation into mediators and then into pairs of $e^+e^-$ or $\mu^+\mu^-$. 
These striking signals are not currently looked at. Furthermore, in these scenarios, it is possible to reconstruct the masses of the DM particle as well as the mediator from the self destruction final states.

\section{Experimental signatures and Model independent searches}\label{sec:pheno}

Each one of the different SDDM scenarios presented above has its own unique signature. Still, these different experimental signatures share certain broad brush features. 
Some of the phenomenology was discussed in the context of the angular momentum model in Section~\ref{sec:signal-rate}.
In this section, we discuss the signal characteristics more broadly. Our aim is not to perform a detailed study of a specific model but rather, to discuss the general properties of the possible signals and suggest a set of relatively model-independent searches that an experiment can perform.

We assume that the couplings between the two sectors are very small.
In that limit, the DM annihilates into mediators that
eventually decay to SM particles rather than into the Standard Model directly. 
The signal characteristics depend on several parameters of the model, and we discuss a few of them below. \\[2pt]

{\bf Event Rate:} The rate of events in a detector is discussed in detailed for a specific model in Section~\ref{sec:signal-rate} but will be repeated here for completeness. The rate depends on the scattering cross section $\sigma_\mathrm{scatter}$ which induces the transition from a long-lived state, $\Psi$, to a short-lived one, $\Psi'$. 
As discussed previously, the rate does not depend strongly on the mediator decay length
$L_V\equiv\gamma_V v_V \tau_V$, with $\tau_V$ the mediator lifetime.
If the mediator is short-lived, $L_V<L_\mathrm{det}$ the scattering takes place inside the detector, but if it is long-lived it can also happen in the Earth, outside the detector. However, as long as the mediator decay length is within the size of the Earth, the signal rate is largely independent of the mediator lifetime. This is because the $1/r^2$ fall-off in the flux emanating from a particular scatter site is compensated by the $r^2$ growth in the amount of volume to scatter on. As a result, up to order one geometric factors
\begin{equation} 
\mbox{Rate} \sim nV_\mathrm{det}  \,n_{\Psi}\langle \sigma v\rangle_\mathrm{scatter} \ ,
\end{equation} 
where $V_\mathrm{det}$ is the volume of the detector, $n$ is the nuclear number density either in the detector or in Earth, depending on whether the mediator is long or short lived. Also, $n_{\chi\bar\chi}$ is the number density of long-lived bound states, $\langle\sigma v\rangle_\mathrm{scatter}$ is the velocity averaged cross section to transition from the long lived DM state to its short lived counterpart. Note that the atomic number of the scattering nucleus may be different for $L_V$ larger or smaller than the detector size, which may affect the coherence factor in the cross section.
\\[-2pt]

{\bf Particle ID:}
The mass, spin, and couplings of the mediator determine its decay branching fractions. A well motivated mediator is a dark photon which couples to the SM via kinetic mixing. In this case, the mediator has a sizable branching fraction to decay to electrons as well as to muons, charged pions or heavy flavor (whenever the mediator is above the relevant thresholds).  A scalar or pseudo-scalar mediator could have a sizable branching fraction to the heaviest particle to which it is kinematically allowed. Di-photons are also an interesting final state, particularly for light (pseudo)scalars.  
\\[-2pt]

\begin{figure}[t]
\includegraphics[height=4.5cm]{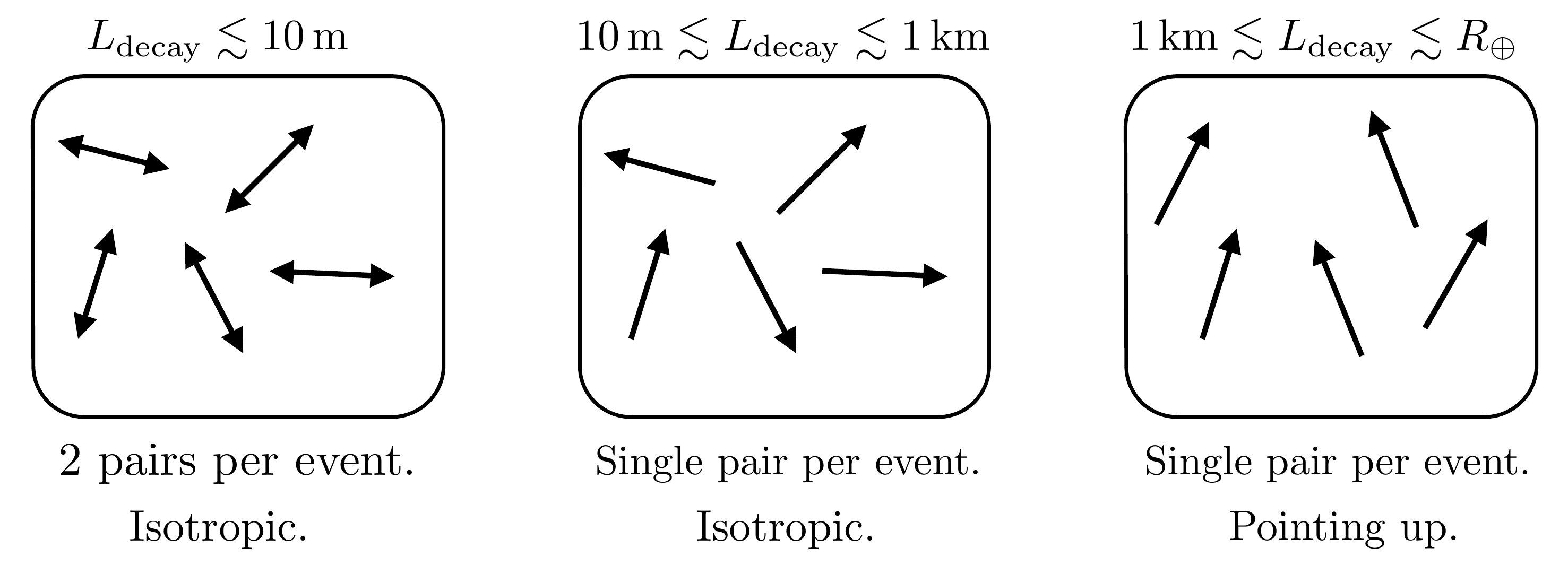}
\caption{A schematic illustration of the expected signals for various decay lengths of the mediator~$V$. In the case of $\Psi'\to2V$ decay, each black arrow represents a pair of SM particles with a total energy of $m_\mathrm{\Psi'}/2$, an invariant mass of $m_V$, and a total momentum pointing in the arrow direction. A double arrow represents two such pairs back-to-back. }
\label{fig:pheno}
\end{figure}

{\bf Event Multiplicity: }
The decay length of the mediators in the lab frame is
\beq
L_\mathrm{V} = c\tau_V \sqrt{\frac{E_V^2}{m_V^2} -1} \ .
\eeq
For $\Psi'\to 2V$ decay, $E_V=m_{\Psi'}/2$ while for $\Psi'\to 3V$ decay, there is a spectrum for $E_V$ between $0$ and $m_{\Psi'}/2$.
If this decay length is shorter than the size of the detector, there can be a sizable fraction of the events in which both mediators decay inside the detector. The signature, in this case, is two simultaneous pairs of leptons with opposite reconstructed velocities (since the decaying bound state is non-relativistic), consistent with a single vertex where the bound state decay took place. This is schematically shown in the left panel of Figure~\ref{fig:pheno}. Such events are expected to have an extremely low background.\\[-2pt]

{\bf Event Kinematics:} Each event consists of decays of on-shell mediators produced via the annihilation $\chi\bar\chi \to 2V$ or $3V$ inside the short-lived bound state $\Psi'$. Consequently, we expect a double or triple lepton pair-production. In the case of $\Psi'\to2V$ decay, each pair has energy which is half of the dark  bound state mass and an invariant mass of $m_V$,
\beq 
\label{eq:kinematics}
E_\mathrm{pair} = \frac{m_{\Psi'}}{2}  \qquad\mbox{and} \qquad   m_\mathrm{pair} = m_V\,. 
\eeq 
\\[-2pt]

{\bf Opening Angle:}
As the mediators in all of our scenarios are produced boosted, their decay products can't be back-to-back. When the decay products are relativistic, the typical opening angle between the, say, $\ell^+$ and $\ell^-$ in each pair  roughly satisfies
\beq
\cos\theta_{\ell^+\ell^-} \sim 1- \frac{8 m_V^2}{m_{\Psi'}^2}  \ ,
\eeq
in the case of $\Psi'\to 2V$ decay.
As long as $m_{V}$ is not too small, detectors with good angular resolution will be able to distinguish the pair from a single energetic particle. On the other hand, as long as the mediator is not too close to half the bound state mass, the pairs will not be back-to-back and the detector will be able to reconstruct the direction of the mediator velocity.\\[-2pt]

{\bf Event Directionality:}
When the decay length of the mediator is much larger than the detector, each event will only consist of a single pair. This region of parameter space can be further categorized based on the directional distribution of the mediator velocity. If the decay length is shorter than the depth of the underground laboratory, the events will be isotropic.  If on the other hand, the decay length is larger than the laboratory depth, the reconstructed velocities would be primarily coming from below. These two possibilities are also shown schematically in Figure~\ref{fig:pheno}.\\[-2pt]

{\bf Associated Signals:}
The transition from the long-lived to the short-lived state could also lead to a visible signal. In this case, if $L_V$ is smaller than the detector size, both the transition and the decay of the mediator(s) can be visible. For example,
in the model described in section~\ref{secSymmetry}, the transition from the long-lived to the short-lived state was associated with the removal of a neutron from a target nucleus which would lead to an additional energy deposit. So long as this process, as well as the decay time of the short live state,  are prompt, the mediator velocity would then point back to the transition point. This class of events can also have very low backgrounds. \\[-2pt]

{\bf Simplified Model Searches:}
Having discussed the signal characteristics, we finally suggest simplified model searches that large detectors can perform. Of course, picking a model and presenting limits in the spirit of Figure~\ref{moneyplot} can certainly be useful, in the sense that a comparison to other searches can be made. However, a more model independent presentation of the limits, which can then be translated to any model, may also be useful. This is common practice in collider searches, where a limit on a particular final state, say a di-lepton resonance, is presented as a limit on a cross section times a branching fraction rather as a limit on couplings. Other examples are searches for simplified models of, say, supersymmetry, where results are shown as limits on the cross section in the stop-neutralino mass plane using a contour or a color scale. 

In the case of SDDM, the signal consists of one or two lepton pairs. The invariant mass of the pairs is set by $m_V$ and the energy is set by $m_{\Psi'}$, see equation~(\ref{eq:kinematics}). A useful way to present results is thus to show the limit on the event rate per unit volume in the $m_V$-$m_{\Psi'}$ plane. To further sophisticate the analysis, this should be repeated for three different assumptions about the decay length $L_V$ which correspond to different angular distributions and multiplicities as shown in Figure~\ref{fig:pheno}, since each of these options can have a different background rate. In the same way that LHC resonances are searched for in a variety of final states, here too, searches can be carried out for pairs of electrons, muons, photons, pions, etc. \\[-2pt]

We conclude that the framework of self-destructing DM has a rich phenomenology. The predicted events can have very particular characteristics which in many cases have very low backgrounds. The background rates and rejection efficiencies can be studied systematically for current and future detectors on a phenomenological basis, varying the model parameters $m_{\Psi'}$, $m_V$ and $\tau_V$ as well as the decay products (electrons, muons, etc).

\section{Conclusions} \label{sec:conc}

The search for DM is currently at the frontiers of high energy physics. In recent years, great efforts have been made in the direct detection of dark matter, aiming at recoil energies around the keV scale or even lower.
This is based on the assumption that the available energy in each dark matter scattering is no larger than its kinetic energy.
In this paper, we have proposed a novel class of DM models called self-destructing Dark Matter (SDDM), in which the scattering of a DM with the detector (or the Earth) induces its decay to SM particles.
The striking new feature of this class of models is the conversion of the entire rest mass of the SDDM to a detectable signal, rather than just its recoil energy. We have demonstrated how large neutrino detectors such as Super-K and DUNE could be at the frontier of the search for SDDM.
Additionally, we presented three concrete realizations of SDDM, all of which are based on the DM being a bound state of some dark interaction, with qualitatively different stabilization mechanisms.  Finally, we briefly described the broad brush features of the expected detector signals in the SDDM scenario.

There are several directions for further study. More detailed studies of the background to SDDM in large neutrino detectors are required, as well as a closer look at its production in the early universe.
Though outside the scope of the current paper, SDDM also has potentially interesting indirect signatures. The most striking one is DM decay in the Sun or in Jupiter. In traditional DM models, the DM may be captured in the core of the Sun, yielding an annihilation signal. In our model, however, the DM can simply ``self-destruct'' in the Sun in a similar manner to Earth. This could yield, {\it e.g.}, gamma ray or $e^+e^-$ signals for indirect detection experiments~\cite{Schuster:2009fc,Feng:2016ijc}.
Additional model dependent indirect signals could arise from DM self destruction induced by DM-DM scattering in the galactic center or in extragalactic sources. 

\section*{ACKNOWLEDGEMENTS}
We thank Kaustubh Agashe, Zackaria Chacko, Andr\'e de Gouv\^ea, Shmuel Nussinov, Surjeet Rajendran, Harikrishnan Ramani, Kate Scholberg, and Yael Shadmi for useful discussions.
This work is supported in part by the DOE grant DE-SC0010143 and in part by
the NSF grant PHY-1719877. Fermilab is operated by Fermi Research Alliance, LLC under Contract No. 
DE-AC02-07CH11359 with the United States Department of Energy.

\appendix

\section{Estimation of Rates in the tunneling Model}
\begin{figure}
\centering
\hspace*{-0.1cm}{\centering 
\includegraphics[height=5.cm]{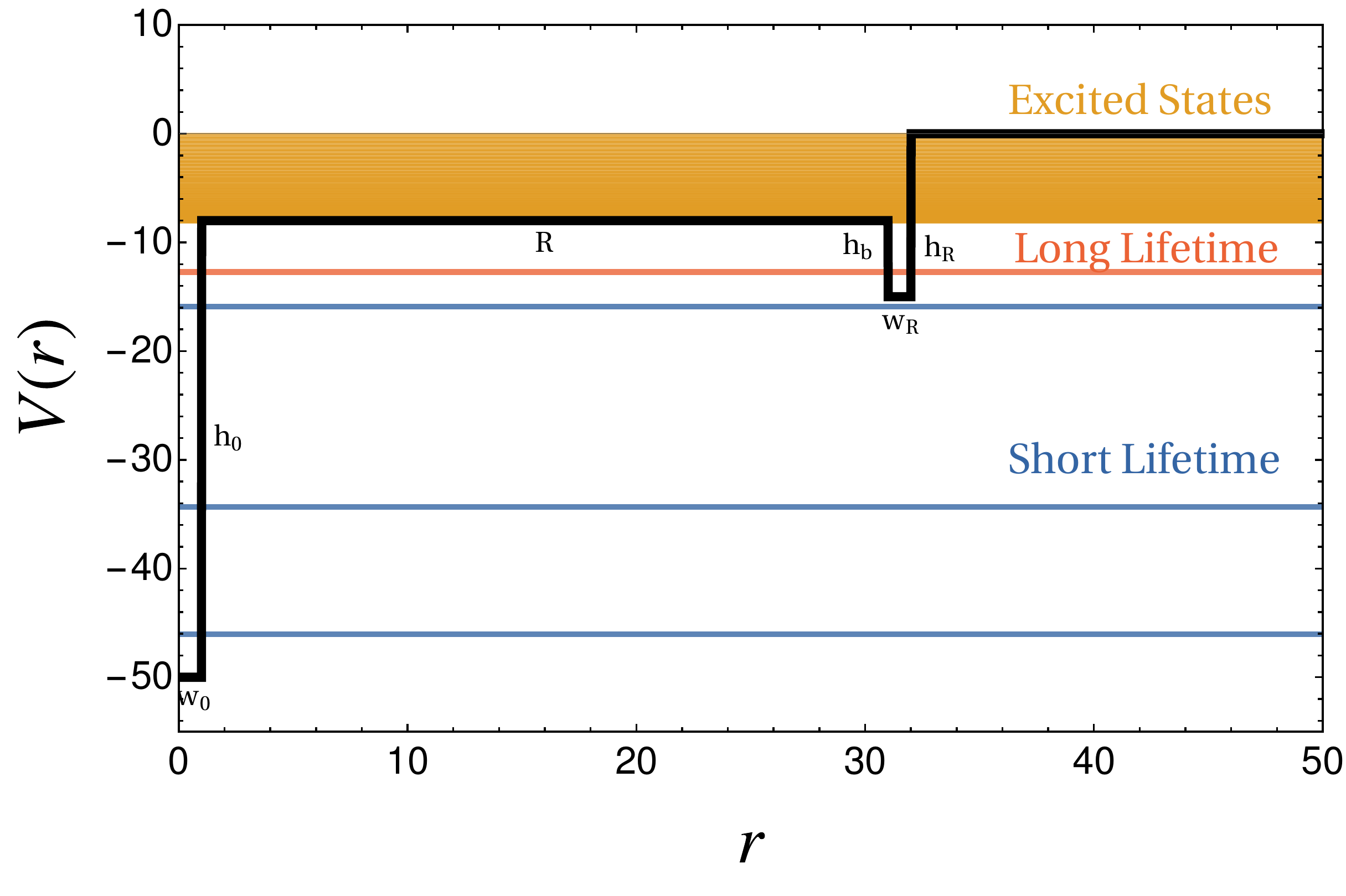}
\hspace*{0.1cm}
\includegraphics[height=5.cm]{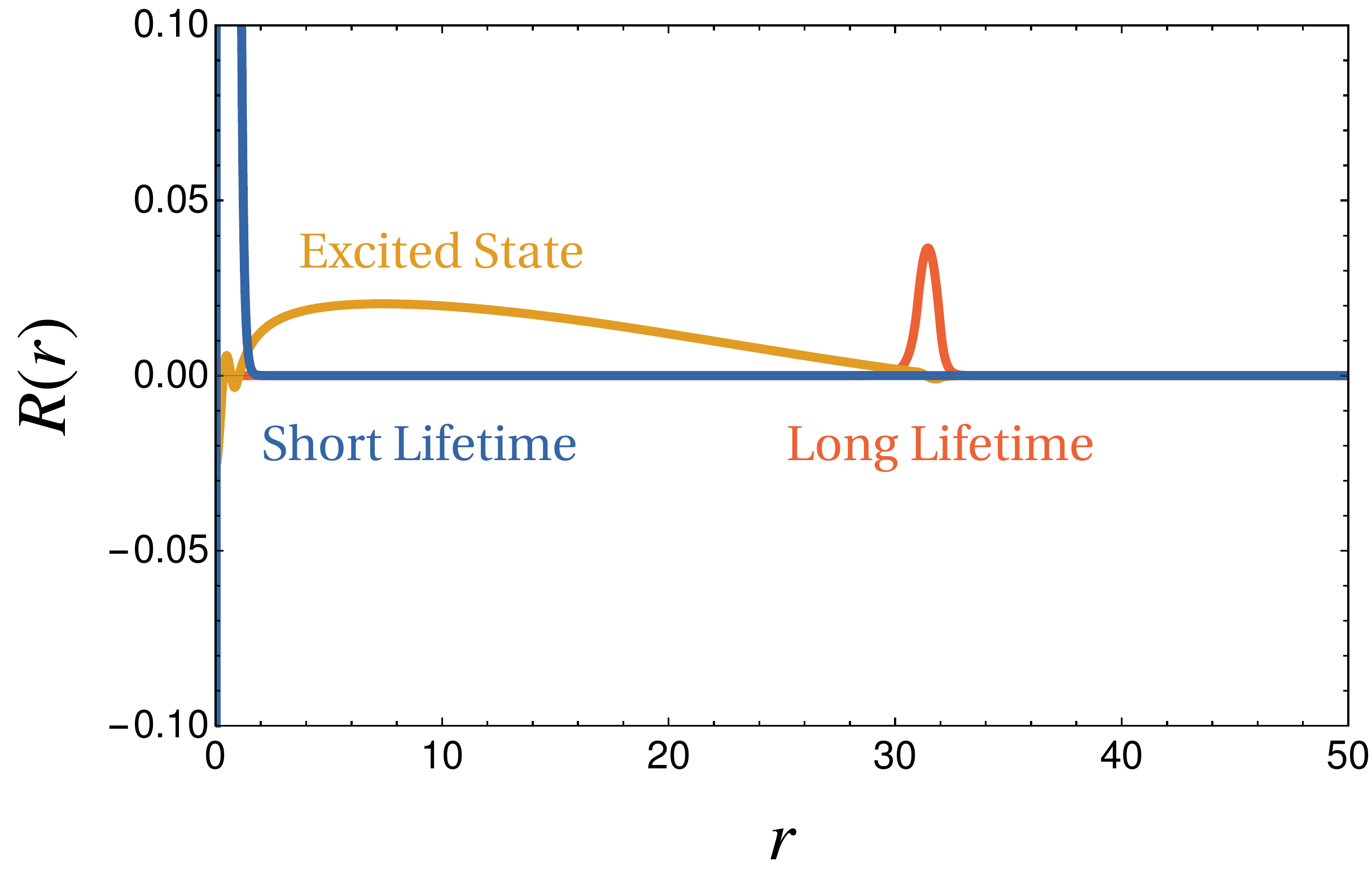}}
\caption{An illustration of tunneling Stabilization. {\it Left}: The potential and energy eigenvalues. {\it Right}: Sample wavefunctions.
The coordinate $r$ is in unit of $m_\chi^{-1}$, the potential energy $V$ is in unit of $m_\chi$ and the bound state wavefunctions are in unit of $m_\chi^{3/2}$.}
\label{fig:ideal}
\end{figure}

To estimate the relevant rates in the tunneling stabilization model discussed in section~\ref{sectunnelling}, we focus on an idealized potential model
 \begin{eqnarray}\label{eq:tpot}
V(r)=\begin{cases}
-h_0 & r<w_0\\
-h_R+h_b & w_0<r<w_0+R\\
-h_R & w_0+R<r<w_0+R+w_R\\
\end{cases}
\end{eqnarray}
The results derived in this section, however are quite generic for more realistic two minima potentials with a wide barrier. The potential Eq.~(\ref{eq:tpot}) is depicted in Fig.~\ref{fig:ideal} (left) for $h_0=50, h_R=15,w_0=w_R=1$ and $h_b=8, R=30$, where $h_0, h_R, h_b$ are in units of $m_\chi$ and $w_0, w_R, R$ are in units of $m_\chi^{-1}$.

We denote the height and width of the barrier by $h_b$ and $R$, respectively. 
The height and width of the potential wells are denoted $h_0, w_0$ and $h_R, w_R$ for the $r=0$ and $r=R$ minima.
The typical wavefunctions of the relevant states are shown in Fig.~\ref{fig:ideal} (right),
and their values near the origin are then,
\begin{eqnarray}
{\left|\Psi_{\text{Long}}(0)\right|}^2\sim \frac{1}{w_R}\,e^{-2\sqrt{h_b}R} \, , \hspace{1cm} {\left|\Psi_{\text{Short}}(0)\right|}^2\sim \frac{1}{w_0} \, , \hspace{1cm} {\left|\Psi_{\text{Excited}}(0)\right|}^2 \sim \frac{1}{R}\ .
\end{eqnarray}
As a result, we have
\begin{eqnarray}
\frac{\tau_{\text{Short}}}{\tau_{\text{Long}}} \sim \frac{\tau_{\text{Excited}}}{\tau_{\text{Long}}}  \sim e^{-2\sqrt{h_b}R} \ .
\end{eqnarray}
For $2\sqrt{h_b}R\sim 40$, the long lifetime states could be cosmologically stable while the short lifetime ones decay in a second. 
The rate for spontaneous emission $\Psi_{\text{Long}}\rightarrow \Psi_{\text{Short}}$ is
\begin{eqnarray}
\Gamma^{SE}_{L\rightarrow S}=\frac{4\alpha_V (\Delta E)^3}{3}\, \left| \int\, d^3r\,\Psi^*_{\text{Short}}(r)\,\vec{r}\,\Psi_{\text{Long}}(r) \right|^2\sim e^{-2\sqrt{h_b}R}\, ,
\end{eqnarray}
and so the long lifetime states are cosmologically stable. In the above estimate, we assume the energy difference $\Delta E$ is large enough for a dark photon to be radiated
on-shell in this model.

\begin{figure}
\centering
\includegraphics[height=6cm]{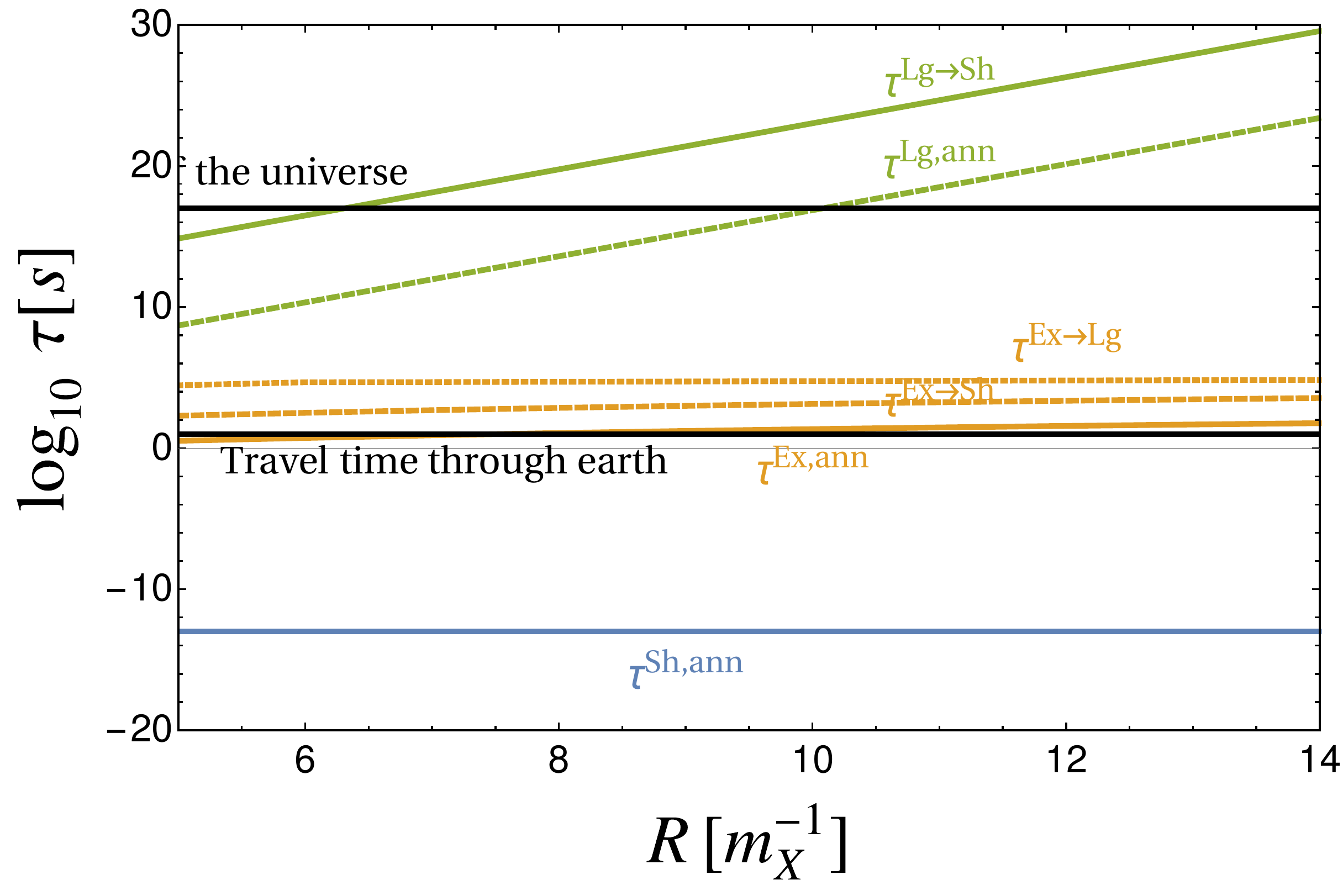}
\caption{Lifetimes for states in the two minima potential as function of the barrier width, for $m_\chi=0.5\text{GeV},\, \alpha_V=10^{-2}$.}
\label{fig:lt}
\end{figure}

The cross section for a long lifetime state to hit the Earth and up scatter to an excited state is given by Eqs.~(\ref{eq:ff})-(\ref{eq:cs}). The form factor $F_D(|\vec{q}\,|)$ is given by integral between $\Psi_{\text{Long}}$ and $\Psi_{\text{Excited}}$ which, given a large enough momentum transfer, is not exponentially suppressed. Once in the excited state, the total rate for spontaneous emission to short lifetime states is
\begin{eqnarray}
\Gamma^{SE}_{E\rightarrow S}=\sum_i\, \frac{4\alpha_V (\Delta E_i)^3}{3}\, \left| \int\, d^3r\,\Psi^*_{\text{Short,i}}(r)\,\vec{r}\,\Psi_{\text{Excited}}(r) \right|^2 \ ,
\end{eqnarray}
which is also free from the exponential suppression. For $N_{\text{Short}}\gg N_{\text{Long}}$, the excited states transit exclusively to short lifetime states. A quantitative analysis of the rates in this model is given in Fig.~\ref{fig:lt} for the simple two minima potential Eq.~(\ref{eq:tpot}), with $m_\chi=0.5\text{GeV},\, \alpha_V=10^{-2}$. As we can see in the plot, the only lifetimes that depend significantly on $R$ are the lifetimes of the long lifetime states to annihilate or spontaneously emit a dark photon and transfer to a short lifetime state. This is because the tunneling suppression for these states depends exponentially on $R$. For $R\gtrsim 12\, m^{-1}_\chi,$ the long lifetime states are cosmologically stable, while the short lifetime states annihilate in roughly a picosecond. Interestingly, for this particular configuration, the excited states tend to annihilate rather than spontaneously emit to either the short or long lifetime states. This is, of course, a completely viable option from the point of view of self-destructing phenomenology.



\begin{thebibliography}{99}

\bibitem{Goodman:1984dc} 
  M.~W.~Goodman and E.~Witten,
  ``Detectability of Certain Dark Matter Candidates,''
  Phys.\ Rev.\ D {\bf 31}, 3059 (1985).

\bibitem{TuckerSmith:2001hy} 
  D.~Tucker-Smith and N.~Weiner,
  ``Inelastic Dark Matter,''
  Phys.\ Rev.\ D {\bf 64}, 043502 (2001)
  [hep-ph/0101138].

\bibitem{Finkbeiner:2007kk} 
  D.~P.~Finkbeiner and N.~Weiner,
  ``Exciting Dark Matter and the INTEGRAL/SPI 511 keV signal,''
  Phys.\ Rev.\ D {\bf 76}, 083519 (2007)
  [astro-ph/0702587].

\bibitem{Chang:2008gd} 
  S.~Chang, G.~D.~Kribs, D.~Tucker-Smith and N.~Weiner,
  Phys.\ Rev.\ D {\bf 79}, 043513 (2009)
  doi:10.1103/PhysRevD.79.043513
  [arXiv:0807.2250 [hep-ph]].

\bibitem{Bramante:2016rdh} 
  J.~Bramante, P.~J.~Fox, G.~D.~Kribs and A.~Martin,
  Phys.\ Rev.\ D {\bf 94}, no. 11, 115026 (2016)
  doi:10.1103/PhysRevD.94.115026
  [arXiv:1608.02662 [hep-ph]].

\bibitem{Chang:2008xa} 
  S.~Chang, A.~Pierce and N.~Weiner,
  Phys.\ Rev.\ D {\bf 79}, 115011 (2009)
  doi:10.1103/PhysRevD.79.115011
  [arXiv:0808.0196 [hep-ph]].

\bibitem{Graham:2010ca} 
  P.~W.~Graham, R.~Harnik, S.~Rajendran and P.~Saraswat,
  ``Exothermic Dark Matter,''
  Phys.\ Rev.\ D {\bf 82}, 063512 (2010)
  [arXiv:1004.0937 [hep-ph]].

\bibitem{Feldstein:2010su} 
  B.~Feldstein, P.~W.~Graham and S.~Rajendran,
  ``Luminous Dark Matter,''
  Phys.\ Rev.\ D {\bf 82}, 075019 (2010)
  [arXiv:1008.1988 [hep-ph]].

\bibitem{Essig:2010ye} 
  R.~Essig, J.~Kaplan, P.~Schuster and N.~Toro,
  ``On the Origin of Light Dark Matter Species,''
  [arXiv:1004.0691 [hep-ph]].

\bibitem{Bai:2009cd} 
  Y.~Bai and P.~J.~Fox,
  ``Resonant Dark Matter,''
  JHEP {\bf 0911}, 052 (2009)
  [arXiv:0909.2900 [hep-ph]].

\bibitem{Pospelov:2008qx} 
  M.~Pospelov and A.~Ritz,
  Phys.\ Rev.\ D {\bf 78}, 055003 (2008)
  doi:10.1103/PhysRevD.78.055003
  [arXiv:0803.2251 [hep-ph]].

\bibitem{An:2012bs} 
  H.~An, M.~Pospelov and J.~Pradler,
  Phys.\ Rev.\ Lett.\  {\bf 109}, 251302 (2012)
  doi:10.1103/PhysRevLett.109.251302
  [arXiv:1209.6358 [hep-ph]].

\bibitem{Chang:2009yt} 
  S.~Chang, A.~Pierce and N.~Weiner,
  JCAP {\bf 1001}, 006 (2010)
  doi:10.1088/1475-7516/2010/01/006
  [arXiv:0908.3192 [hep-ph]].

\bibitem{Chang:2010en} 
  S.~Chang, N.~Weiner and I.~Yavin,
  ``Magnetic Inelastic Dark Matter,''
  Phys.\ Rev.\ D {\bf 82}, 125011 (2010)
  [arXiv:1007.4200 [hep-ph]].

\bibitem{Pospelov:2013nea} 
  M.~Pospelov, N.~Weiner and I.~Yavin,
  Phys.\ Rev.\ D {\bf 89}, no. 5, 055008 (2014)
  [arXiv:1312.1363 [hep-ph]].

\bibitem{Kumar:2011iy} 
  K.~Kumar, A.~Menon and T.~M.~P.~Tait,
  ``Magnetic Fluffy Dark Matter,''
  JHEP {\bf 1202}, 131 (2012)
  [arXiv:1111.2336 [hep-ph]].



\bibitem{Davoudiasl:2010am} 
  H.~Davoudiasl, D.~E.~Morrissey, K.~Sigurdson and S.~Tulin,
  ``Hylogenesis: A Unified Origin for Baryonic Visible Matter and Antibaryonic Dark Matter,''
  Phys.\ Rev.\ Lett.\  {\bf 105}, 211304 (2010)
  [arXiv:1008.2399 [hep-ph]].

\bibitem{Davoudiasl:2011fj} 
  H.~Davoudiasl, D.~E.~Morrissey, K.~Sigurdson and S.~Tulin,
  ``Baryon Destruction by Asymmetric Dark Matter,''
  Phys.\ Rev.\ D {\bf 84}, 096008 (2011)
  [arXiv:1106.4320 [hep-ph]].

\bibitem{Huang:2013xfa} 
  J.~Huang and Y.~Zhao,
  ``Dark Matter Induced Nucleon Decay: Model and Signatures,''
  JHEP {\bf 1402}, 077 (2014)
  [arXiv:1312.0011 [hep-ph]].

\bibitem{Agashe:2014yua} 
  K.~Agashe, Y.~Cui, L.~Necib and J.~Thaler,
  ``(In)direct Detection of Boosted Dark Matter,''
  JCAP {\bf 1410}, no. 10, 062 (2014)
  [arXiv:1405.7370 [hep-ph]].

\bibitem{Berger:2014sqa} 
  J.~Berger, Y.~Cui and Y.~Zhao,
  ``Detecting Boosted Dark Matter from the Sun with Large Volume Neutrino Detectors,''
  JCAP {\bf 1502}, no. 02, 005 (2015)
  [arXiv:1410.2246 [hep-ph]].

\bibitem{Kong:2014mia} 
  K.~Kong, G.~Mohlabeng and J.~C.~Park,
  ``Boosted dark matter signals uplifted with self-interaction,''
  Phys.\ Lett.\ B {\bf 743}, 256 (2015)
  [arXiv:1411.6632 [hep-ph]].

\bibitem{Alhazmi:2016qcs} 
  H.~Alhazmi, K.~Kong, G.~Mohlabeng and J.~C.~Park,
  ``Boosted Dark Matter at the Deep Underground Neutrino Experiment,''
  JHEP {\bf 1704}, 158 (2017)
  [arXiv:1611.09866 [hep-ph]].

\bibitem{Kachulis:2017nci} 
  C.~Kachulis {\it et al.} [Super-Kamiokande Collaboration],
  ``Search for Boosted Dark Matter Interacting With Electrons in Super-Kamiokande,''
  arXiv:1711.05278 [hep-ex].

\bibitem{Sikivie:1983ip} 
  P.~Sikivie,
  ``Experimental Tests of the Invisible Axion,''
  Phys.\ Rev.\ Lett.\  {\bf 51}, 1415 (1983)
  Erratum: [Phys.\ Rev.\ Lett.\  {\bf 52}, 695 (1984)].
  doi:10.1103/PhysRevLett.51.1415, 10.1103/PhysRevLett.52.695.2

\bibitem{Chaudhuri:2014dla} 
  S.~Chaudhuri, P.~W.~Graham, K.~Irwin, J.~Mardon, S.~Rajendran and Y.~Zhao,
  Phys.\ Rev.\ D {\bf 92}, no. 7, 075012 (2015)
  doi:10.1103/PhysRevD.92.075012
  [arXiv:1411.7382 [hep-ph]].

\bibitem{Hochberg:2016sqx} 
  Y.~Hochberg, T.~Lin and K.~M.~Zurek,
  Phys.\ Rev.\ D {\bf 95}, no. 2, 023013 (2017)
  doi:10.1103/PhysRevD.95.023013
  [arXiv:1608.01994 [hep-ph]].

\bibitem{Arvanitaki:2008hq} 
  A.~Arvanitaki, S.~Dimopoulos, S.~Dubovsky, P.~W.~Graham, R.~Harnik and S.~Rajendran,
  Phys.\ Rev.\ D {\bf 79}, 105022 (2009)
  doi:10.1103/PhysRevD.79.105022
  [arXiv:0812.2075 [hep-ph]].

\bibitem{Arvanitaki:2009yb} 
  A.~Arvanitaki, S.~Dimopoulos, S.~Dubovsky, P.~W.~Graham, R.~Harnik and S.~Rajendran,
  Phys.\ Rev.\ D {\bf 80}, 055011 (2009)
  doi:10.1103/PhysRevD.80.055011
  [arXiv:0904.2789 [hep-ph]].

\bibitem{Essig:2013goa} 
  R.~Essig, E.~Kuflik, S.~D.~McDermott, T.~Volansky and K.~M.~Zurek,
  ``Constraining Light Dark Matter with Diffuse X-Ray and Gamma-Ray Observations,''
  JHEP {\bf 1311}, 193 (2013)
  [arXiv:1309.4091 [hep-ph]].

\bibitem{An:2016gad} 
  H.~An, M.~B.~Wise and Y.~Zhang,
  ``Effects of Bound States on Dark Matter Annihilation,''
  Phys.\ Rev.\ D {\bf 93}, no. 11, 115020 (2016)
  [arXiv:1604.01776 [hep-ph]].

\bibitem{Jungman:1995df} 
  G.~Jungman, M.~Kamionkowski and K.~Griest,
  ``Supersymmetric Dark Matter,''
  Phys.\ Rept.\  {\bf 267}, 195 (1996)
  [hep-ph/9506380].

\bibitem{Alexander:2016aln} 
  J.~Alexander {\it et al.},
  ``Dark Sectors 2016 Workshop: Community Report,''
  arXiv:1608.08632 [hep-ph].

\bibitem{Richard:2015aua} 
  E.~Richard {\it et al.} [Super-Kamiokande Collaboration],
  ``Measurements of the atmospheric neutrino flux by Super-Kamiokande: energy spectra, geomagnetic effects, and solar modulation,''
  Phys.\ Rev.\ D {\bf 94}, no. 5, 052001 (2016)
  [arXiv:1510.08127 [hep-ex]].

\bibitem{Andringa:2015tza} 
  S.~Andringa {\it et al.} [SNO+ Collaboration],
  ``Current Status and Future Prospects of the SNO+ Experiment,''
  Adv.\ High Energy Phys.\  {\bf 2016}, 6194250 (2016)
  [arXiv:1508.05759 [physics.ins-det]].

\bibitem{Bellini:2008mr} 
  G.~Bellini {\it et al.} [Borexino Collaboration],
  ``Measurement of the solar 8B neutrino rate with a liquid scintillator target and 3 MeV energy threshold in the Borexino detector,''
  Phys.\ Rev.\ D {\bf 82}, 033006 (2010)
  [arXiv:0808.2868 [astro-ph]].

\bibitem{Acciarri:2015uup} 
  R.~Acciarri {\it et al.} [DUNE Collaboration],
  ``Long-Baseline Neutrino Facility (LBNF) and Deep Underground Neutrino Experiment (DUNE) : Volume 2: The Physics Program for DUNE at LBNF,''
  arXiv:1512.06148 [physics.ins-det].



\bibitem{Wise:2014jva} 
  M.~B.~Wise and Y.~Zhang,
  ``Stable Bound States of Asymmetric Dark Matter,''
  Phys.\ Rev.\ D {\bf 90}, no. 5, 055030 (2014)
  Erratum: [Phys.\ Rev.\ D {\bf 91}, no. 3, 039907 (2015)]
  [arXiv:1407.4121 [hep-ph]].

\bibitem{Madhavacheril:2013cna} 
  M.~S.~Madhavacheril, N.~Sehgal and T.~R.~Slatyer,
  ``Current Dark Matter Annihilation Constraints from CMB and Low-Redshift Data,''
  Phys.\ Rev.\ D {\bf 89}, 103508 (2014)
  [arXiv:1310.3815 [astro-ph.CO]].

\bibitem{Graesser:2011wi} 
  M.~L.~Graesser, I.~M.~Shoemaker and L.~Vecchi,
  ``Asymmetric WIMP Dark Matter,''
  JHEP {\bf 1110}, 110 (2011)
  [arXiv:1103.2771 [hep-ph]].

\bibitem{Gelmini:2013awa} 
  G.~B.~Gelmini, J.~H.~Huh and T.~Rehagen,
  ``Asymmetric dark matter annihilation as a test of non-standard cosmologies,''
  JCAP {\bf 1308}, 003 (2013)
  doi:10.1088/1475-7516/2013/08/003
  [arXiv:1304.3679 [hep-ph]].

\bibitem{Kang:2006yd} 
  J.~Kang, M.~A.~Luty and S.~Nasri,
  ``The Relic abundance of long-lived heavy colored particles,''
  JHEP {\bf 0809}, 086 (2008)
  [hep-ph/0611322].

\bibitem{TheLIGOScientific:2017qsa} 
  B.~P.~Abbott {\it et al.} [LIGO Scientific and Virgo Collaborations],
  ``GW170817: Observation of Gravitational Waves from a Binary Neutron Star Inspiral,''
  Phys.\ Rev.\ Lett.\  {\bf 119}, no. 16, 161101 (2017)
  [arXiv:1710.05832 [gr-qc]].

\bibitem{supercool}
F.~H.~Stillinger ,
``A Topographic View of Supercooled Liquids and Glass Formation,''
Science {\bf 267}, 1935-1939 (1995)

\bibitem{Wise:2014ola} 
  M.~B.~Wise and Y.~Zhang,
  ``Yukawa Bound States of a Large Number of Fermions,''
  JHEP {\bf 1502}, 023 (2015)
  Erratum: [JHEP {\bf 1510}, 165 (2015)]
  [arXiv:1411.1772 [hep-ph]].

\bibitem{Gresham:2017zqi} 
  M.~I.~Gresham, H.~K.~Lou and K.~M.~Zurek,
  ``Nuclear Structure of Bound States of Asymmetric Dark Matter,''
  arXiv:1707.02313 [hep-ph].


\bibitem{Schuster:2009fc} 
  P.~Schuster, N.~Toro, N.~Weiner and I.~Yavin,
  ``High Energy Electron Signals from Dark Matter Annihilation in the Sun,''
  Phys.\ Rev.\ D {\bf 82}, 115012 (2010)
  [arXiv:0910.1839 [hep-ph]].

\bibitem{Feng:2016ijc} 
  J.~L.~Feng, J.~Smolinsky and P.~Tanedo,
  ``Detecting dark matter through dark photons from the Sun: Charged particle signatures,''
  Phys.\ Rev.\ D {\bf 93}, no. 11, 115036 (2016)
  Erratum: [Phys.\ Rev.\ D {\bf 96}, no. 9, 099903 (2017)]
  [arXiv:1602.01465 [hep-ph]].

\end{thebibliography}
\end{document}